\documentclass{article}

    \PassOptionsToPackage{numbers, compress}{natbib}
 \usepackage[main, final]{neurips_2025}

\usepackage[utf8]{inputenc} % allow utf-8 input
\usepackage[T1]{fontenc}    % use 8-bit T1 fonts
\usepackage{url}            % simple URL typesetting
\usepackage{booktabs}       % professional-quality tables
\usepackage{amsfonts}       % blackboard math symbols
\usepackage{nicefrac}       % compact symbols for 1/2, etc.
\usepackage{microtype}      % microtypography
\usepackage{xcolor}         % colors

\usepackage{graphicx} % 
\usepackage{colortbl} % 
\usepackage{multirow}
\usepackage{booktabs} % 
\usepackage{array} % 
\usepackage[page]{appendix}
\usepackage{lineno}
\usepackage{amsmath}
\usepackage{algorithm}
\usepackage{algorithmic}
\usepackage{enumitem}

\definecolor{oursblue}{RGB}{222,240,255}
\definecolor{baselinegray}{RGB}{245,245,245}
\definecolor{timegray}{RGB}{230,230,230}
\usepackage[pagebackref,breaklinks,colorlinks=true,citecolor=blue!60!black]{hyperref}

\makeatletter
\def\thanks#1{\protected@xdef\@thanks{\@thanks
        \protect\footnotetext{#1}}}

\title{Bézier Splatting for Fast and Differentiable Vector Graphics Rendering}

\author{Xi Liu$^{1*}$,~ Chaoyi Zhou$^{1*}$,~ Nanxuan Zhao$^2$,~ Siyu Huang$^1$\thanks{\emph{Corresponding author: Siyu Huang}}\\ \vspace{-1em}\\
$^1$Clemson University, ~~~ $^2$Adobe Research
\\ \vspace{-1em}\\}

\begin{document}

% \twocolumn[{%
% \renewcommand\twocolumn[1][]{#1}%
% \maketitle
% \begin{center}
%     \centering
%     \captionsetup{type=figure}
%     \vspace{-1.5em}
%     \includegraphics[width=1\linewidth]{figures/teaser.pdf}
%     \vspace{-2em}
%     \captionof{figure}{This work proposes \textbf{Bézier splatting}, a new differentiable vector graphic (VG) representation that enables fast and high-fidelity vectorization of high-resolution images, achieving the outstanding rendering performance, \emph{20$\times$} forward computation speedup, \emph{150$\times$} backward computation speedup, and over \emph{10$\times$} total training speedup (tested on a NVIDIA A100 GPU for open curve) in comparison with the state-of-the-art differentiable VG rasterization methods. DiffVG \cite{Li:2020:DVG} and Bézier splatting support both open and closed curves, while LIVE \cite{xu2022live} supports closed curves only. Zoom in for a better view of details.}
%     \label{fig:teaser}
% \end{center}%
% }]

\maketitle

\begin{figure}[ht]
\vspace{-16pt}
\begin{center}
    \includegraphics[width=1\linewidth]{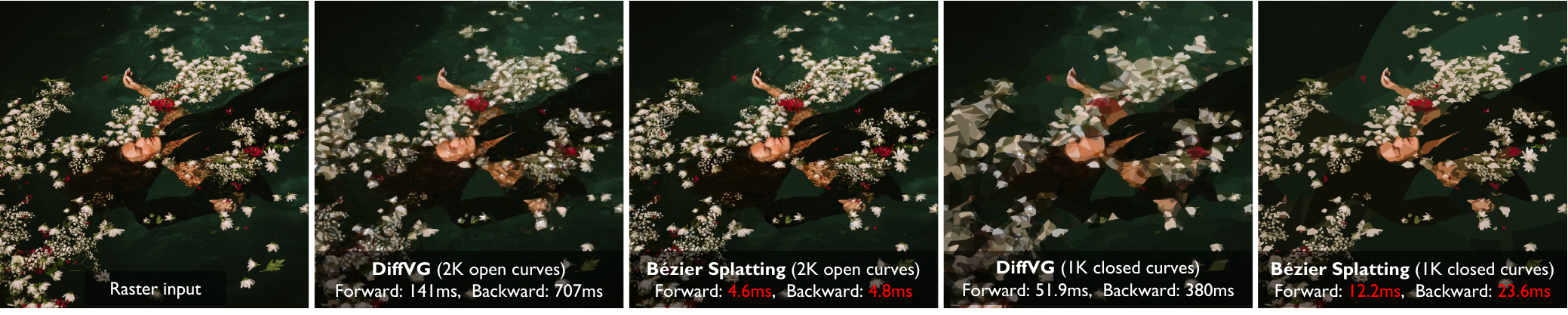}
\end{center}
% \vspace{-7pt}
\vspace{-1em}
% enables fast and high-fidelity vectorization of high-resolution images, 
% , and over \emph{10$\times$} total training speedup
% impressive rendering quality, \emph{20$\times$} faster in forward pass and \emph{150$\times$} faster in backward pass
\caption{This work proposes \textit{Bézier Splatting}, a new differentiable vector graphics (VGs) renderer that achieves an order-of-magnitude computational speedup in comparison with the state-of-the-art method DiffVG \cite{Li:2020:DVG} (tested on a NVIDIA RTX 4090 GPU).
% and Bézier Splatting support both open and closed curves, while LIVE \cite{xu2022live} supports closed curves only. Zoom in for a better view of details.
}
% \vspace{-2pt}
\label{fig:teaser}
\end{figure}

% \begin{center}
%     \centering
%     \includegraphics[width=1\linewidth]{figures/teaser.pdf}
%     \vspace{-2em}
%     \captionof{figure}{This work proposes \textbf{Bézier splatting}, a new differentiable vector graphic (VG) representation that enables fast and high-fidelity vectorization of high-resolution images, achieving the outstanding rendering performance, \emph{20$\times$} forward computation speedup, \emph{150$\times$} backward computation speedup, and over \emph{10$\times$} total training speedup (tested on a NVIDIA A100 GPU for open curve) in comparison with the state-of-the-art differentiable VG rasterization methods. DiffVG \cite{Li:2020:DVG} and Bézier splatting support both open and closed curves, while LIVE \cite{xu2022live} supports closed curves only. Zoom in for a better view of details.}
%     \label{fig:teaser}
% \end{center}

% \vspace{-2em}

\begin{abstract}
Differentiable vector graphics (VGs) are widely used in image vectorization and vector synthesis, while existing representations are costly to optimize and struggle to achieve high-quality rendering results for high-resolution images. This work introduces a new differentiable VG representation, dubbed Bézier Splatting, that enables fast yet high-fidelity VG rasterization. Bézier Splatting samples 2D Gaussians along Bézier curves, which naturally provide positional gradients at object boundaries. Thanks to the efficient splatting-based differentiable rasterizer, Bézier Splatting achieves 30× and 150× faster per forward and backward rasterization step for open curves compared to DiffVG. Additionally, we introduce an adaptive pruning and densification strategy that dynamically adjusts the spatial distribution of curves to escape local minima, further improving VG quality. Furthermore, our new VG representation supports conversion to standard XML-based SVG format, enhancing interoperability with existing VG tools and pipelines. Experimental results show that Bézier Splatting significantly outperforms existing methods with better visual fidelity and significant optimization speedup. 
The project page is \href{https://xiliu8006.github.io/Bezier_splatting_project/}{\color{red!60!red} xiliu8006.github.io/Bezier\_splatting\_project}.
\end{abstract}    
\vspace{-1em}

\section{Introduction}
\label{sec:intro}
\vspace{-.5em}

Vector graphics (VGs) represent images through parametric primitives such as points, curves, and shapes. Unlike raster images, they enable structured representations, lossless resizing, compact storage, and precise content editing, making them crucial for various applications such as user interfaces and animation. 

Recently, differentiable VG rasterization gains significant attention, as it allows raster-based algorithms to edit or synthesize VGs through gradient-based optimization. DiffVG \cite{Li:2020:DVG} is the first differentiable VG framework that leverages the anti-aliasing algorithm to differentiate the vector curves that are inherently discontinuous in pixel space. However, it suffers from slow training and low-fidelity rendering, particularly for high-resolution images. LIVE \cite{xu2022live} further introduces a layer-wise coarse-to-fine strategy to improve quality and topology. However, it is highly computationally expensive, requiring 5 hours to vectorize a 2K-resolution image. Additionally, learning-based methods such as Im2Vec \cite{reddy2021im2vec} train neural networks to map pixels to VGs, but they are restricted to simple graphics and struggle with out-of-domain generalization. Towards scalable applications of VGs, these challenges highlight the need for a differentiable VG method with better efficiency, fidelity, and generalization capability. 

This work presents \emph{Bézier Splatting}, a new differentiable VG representation that optimizes Bézier curves through Gaussian splatting-based rasterization. We sample 2D Gaussian points along Bézier curves and their interior regions, then leverage the Gaussian Splatting framework \cite{kerbl3Dgaussians} for efficient curve rasterization. Unlike DiffVG \cite{Li:2020:DVG} which requires computationally intensive boundary sampling and gradient computation, 2D Gaussians inherently provide direct positional gradients at object boundaries through its differentiable Gaussian formulation, enabling over \emph{150$\times$} faster backward computation over DiffVG for open curves. We further introduce an adaptive pruning and densification strategy that adaptively removes redundant curves while adding new ones to necessary regions during optimization. It helps the optimization process escape the local minima of current spatial distributions of curves, formulating a ``global receptive field'' for further improving the VG optimization process. As shown in Fig. \ref{fig:teaser}, Bézier Splatting outperforms the state-of-the-art differentiable VG rasterizer DiffVG \cite{Li:2020:DVG} for both open and closed curves in terms of efficiency and rendering quality.
% The open curve training takes only $2.9$ minutes, representing an over \emph{10}$\times$ training acceleration compared to DiffVG \cite{Li:2020:DVG} (43 minutes). 
% Bézier Splatting presents better image details and captures fine-grained texture information,
% for instance, the richer facial details.
% While LIVE \cite{Du:2023:IVE} enhances details to some extent, its layer-wise optimization strategy results in artifacts like non-smooth shape boundaries on the sky as well as significantly more computing cost, whereas Bézier Splatting enables 88$\times$ training acceleration. 
We summarize the contribution of this work as follows.
\begin{itemize}\setlength{\itemsep}{3pt}
    \item We propose a novel differentiable vector graphic representation, Bézier Splatting, which achieves an order-of-magnitude computational speedup while producing high-quality rendering results. 
    \item We present an adaptive pruning and densification strategy to improve the optimization process of Bézier curves by escaping the local minima of the spatial distributions of curves.
    \item Extensive experiments demonstrate that Bézier Splatting outperforms existing differentiable VG rendering methods in efficiency and visual quality.
\end{itemize}

\vspace{-.5em}
\section{Related works}
% \vspace{-.5em}

\label{sec:related_works}

\vspace{-.5em}
\subsection{Vectorization and Rasterization}
\vspace{-.5em}

Image vectorization and vector graphic (VG) rasterization are important research topics in computer graphics and computer vision. DiffVG \cite{Li:2020:DVG} introduces the first differentiable VG rasterization framework, laying a foundation for optimizing and generating vectorized representations through gradient-based methods. It expands the applicability of VGs to a broader range of tasks including image vectorization, text-to-VG generation, and painterly rendering. Based on DiffVG \cite{Li:2020:DVG}, LIVE \cite{xu2022live} and O\&R \cite{hirschorn2024optimize} further proposes a layer-wise path initialization strategy, vectorizing raster images into compact and semantically consistent VG representations while preserving image topology. Du~\emph{et al.}~\cite{Du:2023:IVE} use linear gradients to fill the regional colors, and Chen \emph{et al.}~\cite{Chen_2024_CVPR} propose a specific implicit neural representation to model regional color distributions, enhancing the color representation within closed Bézier curves. However, these DiffVG-based VG rasterization approaches \cite{Li:2020:DVG, xu2022live, Du:2023:IVE, Chen_2024_CVPR} suffer from slow optimization, often requiring several hours to process a 2K-resolution image with 1,024 curves. 

With the advancement of deep learning, another line of approaches directly learns deep neural networks for vector synthesis.
Lopes~\emph{et al.}~\cite{lopes2019learned} combine image-based encoder and VG decoder to generate fonts. Img2Vec \cite{reddy2021im2vec} integrates the encoder with recurrent neural networks (RNNs) by leveraging sequential modeling for structured vector synthesis. SVGFormer \cite{Cao_2023_CVPR} further adopts a Transformer-based architecture \cite{vaswani2017attention} to improve the capacity for representing complex geometric structures. More recently, diffusion models \cite{ho2020denoising} have been applied to text-to-VG synthesis \cite{zhang2024text, jain2023vectorfusion, xing2024svgdreamer, xing2023diffsketcher}. However, existing learning-based VG synthesis approaches are limited to generating simple graphics, and struggle with out-of-domain data. In this work, we propose a novel VG representation dubbed Bézier Splatting that achieves high-fidelity VG rendering within minutes of optimization for high-resolution images.

\vspace{-.5em}
\subsection{Gaussian Splatting} 
\vspace{-.5em}

3D Gaussian Splatting \cite{kerbl3Dgaussians, zwicker2001ewa} emerges as a promising approach for novel view synthesis (NVS) and attracts significant attention from the community. Its explicit 3D Gaussian-based volumetric representation enables high-fidelity 3D reconstruction, while the differentiable tile-based rasterization pipeline ensures real-time and high-quality rendering. It has been applied to various domains and tasks, such as 4D modeling \cite{luiten2023dynamic,Wu_2024_CVPR} and 3D scene generation \cite{yi2023gaussiandreamer, tang2024dreamgaussian}.
Several works further improve the Gaussian primitives for better representation quality. SuGaR \cite{guedon2023sugar} proposes to approximate 3D Gaussians with 2D Gaussians for enhanced surface reconstruction. 2D Gaussian Splatting \cite{Huang2DGS2024} directly adopts 2D Gaussians for 3D reconstruction for simplified optimization and improved geometric fidelity.
% without requiring additional mesh refinement. 
TetSphere splatting\cite{guo2024tetsphere} further employs tetrahedral meshes as the geometric primitives to achieve high-quality geometry. 
Particularly for 2D image representation, GaussianImage \cite{zhang2024gaussianimage} adopts 2D Gaussians \cite{Huang2DGS2024} for efficient image representation, achieving a compact and expressive alternative to rasters or implicit representations \cite{sitzmann2020implicit,muller2022instant}. 
Image-GS \cite{zhang2024imagegscontentadaptiveimagerepresentation} further enhances it through a content-adaptive compression approach. This work proposes to integrate Gaussian splatting with vector representations for differentiable VG rasterization. By sampling 2D Gaussians on Bézier curves and rasterizing through the efficient Gaussian splatting pipeline, the proposed Bézier Splatting representation enables fast yet high-quality VG rendering, even for high-resolution images with complex structures.

\vspace{-.5em}
\section{Method}
\vspace{-.5em}

\subsection{Overall}
\vspace{-.5em}

Given a raster image, our goal is to efficiently vectorize it into a VG representation that closely resembles the input while preserving the details. Existing methods, including DiffVG \cite{Li:2020:DVG} and its following work \cite{xu2022live, Chen_2024_CVPR}, incur substantial computational costs due to the pixel color accumulation computation. Specifically, DiffVG first constructs a bounding volume hierarchy (BVH) tree to determine the curves that intersect with each individual pixel, then solves equations to precisely determine whether a pixel lies within a region and to compute the inward or outward gradients at boundary points.

\begin{figure*}[t]
\centering
% \vspace{-.5em}
\includegraphics[width=1\linewidth]{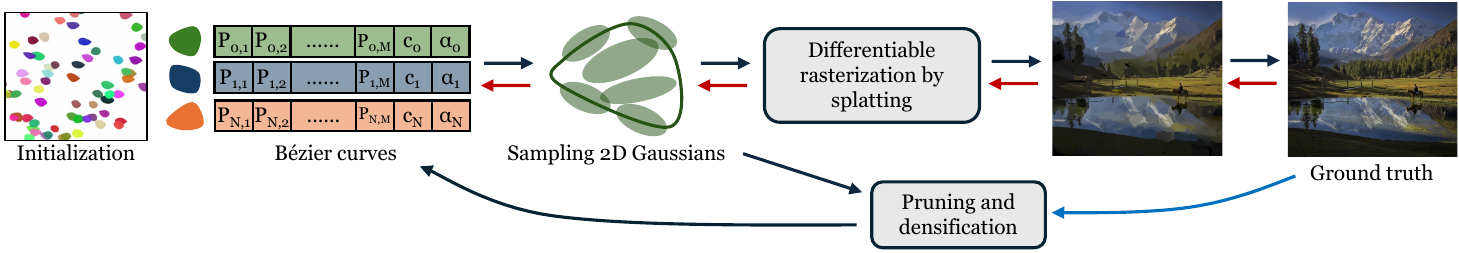}
\caption{An illustration of the algorithm flow of Bézier Splatting. It begins by randomly initializing Bézier curves and uniformly sampling Gaussians points along them. These Gaussians are then rasterized into the image, enabling gradient-based computation to optimize parameters of both Bézier curves and Gaussians. Curves with negligible opacity or extremely small shapes are removed, while new curves are adaptively added into areas with high reconstruction error, ensuring curves are placed in areas requiring finer details. 
% rather than well reconstructed ones.
\textcolor{black}{$\rightarrow$} {forward}, 
\textcolor{red}{$\leftarrow$} backpropagation,
\textcolor{blue}{$\leftarrow$} error map.
}
\label{fig:framework}
\vspace{-.5em}
\end{figure*}

To overcome this inefficiency, this work proposes a novel VG representation, Bézier Splatting, which is inspired by the high computational efficiency and expressive fitting capacity of Gaussian splatting \cite{zhang2024gaussianimage} for rasterization. Bézier Splatting \textit{samples 2D Gaussian points along Bézier curves and their interior regions}, then \textit{leverages
the Gaussian splatting method for efficient rasterization}, enjoying the following advantages:
\begin{itemize}\setlength{\itemsep}{3pt}
    \item This design significantly accelerates the forward and backward pass of Bézier curve rasterization, achieving an order-of-magnitude speedup without specialized optimization techniques;
    \item The 2D Gaussian representation inherently provides direct position gradients for object boundaries, eliminating the need for additional computations such as boundary sampling and gradient derivation via the Reynolds transport theorem in DiffVG \cite{Li:2020:DVG};
    \item It supports richer texture representation by allowing properties such as spatially varying opacity and width, facilitating complex effects such as linear-gradient color transitions within a single curve.
\end{itemize}
To further improve the fidelity and expressiveness of Bézier Splatting, we introduce a pruning and densification approach (Sec. \ref{sec:optimization}), which dynamically removes redundant Bézier curves while adaptively adding necessary curves in regions that have high reconstruction error. Fig. \ref{fig:framework} illustrates the algorithm flow of Bézier Splatting. More details are discussed in the following sections.

\vspace{-.5em}
\subsection{Primitives of Bézier Splatting}
\vspace{-.5em}

\noindent\textbf{Bézier curves.} We adopt Bézier curves as the parametric primitives of VGs. The representation includes \( N \) Bézier curves with a degree of \( M \), as:  
\begin{equation}
\mathcal{B}_i(t) = \sum_{j=0}^{M} B_j^M(t) P^{(i)}_{j}, \; t \in [0,1], \; i \in \{1, \dots, N\},
\label{eq:bezier}
\end{equation}
where $t$ is a normalized position on the curve, \( P^{(i)}_{j} \) represents the \( j \)-th control point of the \( i \)-th Bézier curve, and \( B_j^M(t) \) is the Bernstein polynomial of degree \( M \), given by:  
\begin{equation}
B_j^M(t) = \binom{M}{j} (1 - t)^{M-j} t^j.
\label{eq:bezier1}
\end{equation}
Each Bézier curve \( \mathcal{B}_i(t) \) is associated with an RGB color parameter \( c_i \in \mathbb{R}^3 \), and an opacity parameter \( o_i \in [0,1] \) that defines the transparency of the curve.

To formulate an open curve, we follow DiffVG \cite{Li:2020:DVG} to adopt three sequentially connected Bézier curves with two control points on each. An open curve requires an additional width parameter to define the stroke thickness.
For a closed curve, we adopt two connected Bézier curves, enabling a more efficient sampling in enclosed regions. For closed curves, color filling is applied. The connected Bézier curves in either an open or closed curve share the color parameters, but they have separate opacity parameters to better model the opacity changes along the curves or within closed areas for enriching the texture representation capacity. 

\noindent\textbf{2D Gaussians on curves.}  
Our Bézier Splatting novelly associates 2D Gaussians with each Bézier curve. The standard formulation of 2DGS \cite{Huang2DGS2024} parameterizes each 2D Gaussian by position, color, rotation, scale, opacity, and depth.
However, to ensure a compact and differentiable representation, these parameters of Gaussians in our Bézier Splatting are inherited from the corresponding control points. We discuss more details of the sampling of Gaussians, rasterization process, and backaward computation in the following sections.

\vspace{-.5em}
\subsection{Sampling Gaussians on Bézier Curves}
\vspace{-.5em}

This work introduces a fast differentiable VG rasterizer based on Gaussian splatting, allowing gradients from raster images to be backpropagated to the 2D Gaussians, then further backpropagated to the Bézier curves through a differentiable sampling strategy, resulting in a highly efficient optimization of the Bézier curves.

Specifically, for each Bézier curve $\mathcal{B}_i(t)$, we uniformly sample $K$ points along it based on Eq. \ref{eq:bezier}. The sampled point set $\mathbf{b}_i$ is: 
\begin{equation}
\mathbf{b}_i = \big[ \mathcal{B}_i(t_0), \mathcal{B}_i(t_1), \dots, \mathcal{B}_i(t_{K-1}) \big],
\label{eq:sampling}
\end{equation}
where $t_k$ is uniformly sampled from \([0,1]\).

\noindent\textbf{Sampling 2D Gaussians on open curves.} We represent an open curve by using $3$ sequential Bézier curves $\mathcal{B}_i(t)$ with a degree of $4$ to form a single continuous stroke, by following the same setting as DiffVG \cite{Li:2020:DVG}. The stroke consists of 10 control points, as the end point of each Bézier curve serves as the start point of the next. To ensure that the final rendering result generates a continuous stroke with consistent width and color, we calculate the x-direction scale of a Gaussian point by the distance between neighboring points by following Eq. \ref{eq:sigma}, while the y-direction scale is a learnable parameter that represents the stroke width. 

For both closed and open curves, the depth $d$ of a Gaussian is assigned based on the area of curves, ensuring smaller curves are not occluded by larger ones. It prevents them from being ignored during optimization, as the gradient would not count Gaussians when the accumulated opacity exceeds 1.

\noindent\textbf{Sampling 2D Gaussians on closed curves.} Achieving accurate color filling is non-trivial for closed curves. A straightforward approach involves uniformly sampling a large number of points, identifying intersecting curves, solving equations to determine which curves cover them, and then computing scaling factors based on the sampled points. However, this process is computationally expensive, making it inefficient for curve rasterization and optimization.

% To address this limitation, we propose a new structure named \emph{paired Bézier curve structure} that facilitates an efficient closed area sampling between two Bézier curves of a closed curve. Specifically, the two Bézier curves \( \mathcal{B}_1(t) \) and \( \mathcal{B}_{R+1}(t) \) share the same start and end points, and $R$ intermediate Bézier curves are generated by interpolating the control points between the two Bézier curves $P^{(0)}_j$ and $P^{(R+1)}_j$as:
% \begin{equation}
% P^{(k)}_j = (1 - t_k) P^{(0)}_j + t_k P^{(R+1)}_j, \; k = 1, \dots, R,
% \end{equation}

To address this limitation, we propose a new structure named \emph{paired Bézier curve structure}, which supports two groups of Bézier curves with equal cardinality and arbitrary degrees, forming a closed region. This structure enables efficient and flexible sampling within the enclosed region between the two curve groups. Specifically, the two Bézier curves \( \mathcal{B}_1(t) \) and \( \mathcal{B}_{R+1}(t) \) share the same start and end points, forming the inside curves and boundaries of a closed region. A total of \( R \) intermediate Bézier curves are generated by linearly interpolating the corresponding control points \( P^{(0)}_j \) and \( P^{(R+1)}_j \) as:
\begin{equation}
P^{(k)}_j = (1 - t_k) P^{(0)}_j + t_k P^{(R+1)}_j, \quad k = 1, \dots, R,
\end{equation}
where \( t_k \in [0,1] \) are sampled from a normalized cumulative distribution function (CDF). The interpolated control points \( \{ P^{(k)}_j \} \) define intermediate curves that form a dense strip between the two boundaries. Note that this paired structure can naturally extend to cases where each curve is composed of multiple connected Bézier segments, provided that all segments have the same number of control points, similar to the representation used in DiffVG~\cite{Li:2020:DVG}.

This non-uniform sampling ensures that points near the boundary curves have small scales, mitigating the influence of interior Gaussians to the exterior of closed area. The interpolated Bézier curves are:
\begin{equation}
\mathcal{B}_k^{\text{interp}}(t) = \sum_{j=0}^{M} B_j^M(t) P^{(k)}_j, \quad k = 1, \dots, R.
\end{equation}
Then, 2D Gaussians are sampled on these interpolated curves, by following the same procedure as open curves (Eq. \ref{eq:sampling}), resulting in a structured and efficient point sampling within the enclosed region. Furthermore, to mitigate artifacts caused by non-convex curve shapes during interpolation, we incorporate the Xing loss from LIVE \cite{xu2022live}, which enforces convexity constraints on curve shapes, to improve the stability of the interpolation process.
The full set of sampled points on the $i$-th Bézier curve is:
\begin{equation}
\mathbf{X} = \big[ \mathbf{b}_0, \mathbf{b}_1, \dots, \mathbf{b}_{R+1} \big] \in \mathbb{R}^{{(R+2)} \times K \times 2}.
\end{equation}

% Let $\mathbf{X}_{r,k}$ represents the 2D position of the $k$-th sampled point on the $r$-th interpolated curve, the x-direction scale $\sigma_x(r,k)$ of a 2D Gaussian is calculated by the pairwise distance between consecutive points along the curve, and y-direction scale $\sigma_y(r,k)$ is calculated by the pairwise distance between corresponding points on adjacent curves:
% \begin{equation} \label{eq:sigma}
% \begin{aligned}
% \sigma_x(r,k) &= \| \mathbf{X}_{r,k+1} - \mathbf{X}_{r,k} \|_2 / \rho, \\
% \sigma_y(r,k) &= \| \mathbf{X}_{r+1,k} - \mathbf{X}_{r,k} \|_2 / \rho.
% \end{aligned}
% \end{equation}
% $\rho$ is a constant that determines the density of 2D Gaussian points.
% The rotation $\theta$ of each Gaussian point is calculated by its left and right neighbors, as \begin{equation} \theta_{r,k} = \operatorname{atan2}\left( y_{r,k+1} - y_{r,k-1}, x_{r,k+1} - x_{r,k-1} \right) \label{eq:rotation} \end{equation}
% For points on boundaries, the rotation is set to align with the nearest neighbor.

Let $\mathbf{X}_{r,k}$ denote the 2D position of the $k$-th sampled point on the $r$-th interpolated curve.
To construct anisotropic 2D Gaussian primitives aligned with the local geometry, we define the spatial scales as follows.
The \emph{x-direction} of each 2D Gaussian is defined to align with the local tangent direction of the curve (i.e., along the curve),
while the \emph{y-direction} is perpendicular to it (i.e., across adjacent curves).

The scale $\sigma_x(r,k)$ is computed from the Euclidean distance between consecutive points along the curve, and $\sigma_y(r,k)$ is computed from the distance between corresponding points on adjacent curves:
\begin{equation} \label{eq:sigma}
\begin{aligned}
\sigma_x(r,k) &= | \mathbf{X}_{r,k+1} - \mathbf{X}_{r,k} |_2 / \rho, \\
\sigma_y(r,k) &= | \mathbf{X}_{r+1,k} - \mathbf{X}_{r,k} |_2 / \rho.
\end{aligned}
\end{equation}
Here, $\rho$ is a global constant that controls the overall density and overlap of the Gaussians.
The rotation $\theta_{r,k}$ of each Gaussian is defined by the angle of the local tangent vector, estimated from neighboring points as
\begin{equation} \label{eq:rotation}
\theta_{r,k} = \operatorname{atan2}\left( y_{r,k+1} - y_{r,k-1}, x_{r,k+1} - x_{r,k-1} \right).
\end{equation}
For boundary points, the rotation is set to align with the nearest available neighbor.

\vspace{-.5em}
\subsection{Splatting-based Differentiable Rasterization}
\vspace{-.5em}

\label{sec:rasterizer}
Once all Gaussians on Bézier curves are sampled, the rasterization process follows the Gaussian splatting pipeline \cite{kerbl3Dgaussians}, which is very fast in both forward and backward computation. Different from GaussianImage \cite{zhang2024gaussianimage}, we use $\alpha$-blending for pixel rendering instead of the accumulation-based blending. 
Accumulation-based blending calculates the pixel value based on all overlapping Gaussians, such that it conflicts with the rendering principles of VGs, where occlusion plays a crucial role in defining vector structures.
$\alpha$-blending ensures proper occlusion handling, as it allows foreground elements to contribute more to the final pixel value. 
The rendering of a pixel is:
\begin{equation}
    C_n = \sum_{i \in M}\mathbf{c}_i\alpha_i \prod_{j=1}^{i-1}(1 - \alpha_j).
\end{equation} 
$C_n$ is the $n$-th pixel color, $\mathbf{c}_i$ is the color of corresponding Gaussians, $\alpha_i$ is computed by the projected 2D covariance $\Sigma_i$:
\begin{equation}
\alpha_i = o_i \exp^{-\sigma_i}, \quad 
\sigma_i = \frac{1}{2} \boldsymbol{d}_n^T \boldsymbol{\Sigma}^{-1}_{i} \boldsymbol{d}_n.
\label{eq:rasterization2}
\end{equation}
where $d_n$ is the distance between the pixel and Gaussian center, and $\Sigma_i$ can be modeled by $\theta_i$, $\sigma_x^i$ and $\sigma_y^i$ as:
\begin{equation}
\boldsymbol{\Sigma}_i = (\boldsymbol{R}_i \boldsymbol{S}_i)(\boldsymbol{R}_i \boldsymbol{S}_i)^T.
\label{eq:RSSR}
\end{equation}
\begin{equation}
\boldsymbol{R}_i =
\begin{bmatrix}
\cos(\theta_i) & -\sin(\theta_i) \\
\sin(\theta_i) & \cos(\theta_i)
\end{bmatrix},
\quad
\boldsymbol{S}_i =
\begin{bmatrix}
\sigma_{x}^i & 0 \\
0 & \sigma_{y}^i
\end{bmatrix}.
\label{eq:rot_scale}
\end{equation}

\noindent\textbf{Discussion on method efficiency.}
The rasterization process of Bézier Splatting is very efficient, since we directly sample 2D Gaussians from Bézier curves in a differentiable manner, then splat them to the 2D plane. The rasterization pipeline remains end-to-end differentiable while being highly optimized for parallel computation and large-scale matrix operations.
As a result, Bézier Splatting is a highly efficient and fully differentiable VG representation, which ensures both fast rendering and gradient-based optimization. It not only preserves the flexibility of VGs but also allows seamless integration into deep learning frameworks, making it well-suited for tasks requiring high-quality and editable vector representations.

\vspace{-.5em}
\subsection{Optimization}
\vspace{-.5em}
\label{sec:optimization}

\noindent\textbf{Training objective.}
Given a raster image $\mathcal{I} \in \mathbb{R}^{H \times W \times 3}$, the goal of image vectorization is to vectorize the image into Bézier curves while ensuring a high-fidelity reconstruction. We first randomly generate a set of Bézier curves to lay on the canvas, then employ the differentiable rasterizer discussed in Sec. \ref{sec:rasterizer} to render a raster image $\hat{\mathcal{I}} \in \mathbb{R}^{H \times W \times 3}$. The Bézier curves can then be optimized through any gradient-based loss functions. 
In this work, we formulate the optimization objective as minimizing the loss between  $\mathcal{I}$ and $\hat{\mathcal{I}}$ while enforcing the curves to be convex. Therefore, we only adopt an $L_2$ loss and a Xing loss $L_{\text{Xing}}$  \cite{xu2022live}, as:
\begin{equation}
L = \lambda_1 \|\hat{\mathcal{I}} - \mathcal{I} \|_2^2 +  \lambda_2 L_{\text{Xing}}
\end{equation}
where $\lambda_1$ and $\lambda_2$ are hyperparameters that trade off the two loss functions.

\noindent\textbf{Adaptive curve pruning and densification.}
The gradients of 2D Gaussians are influenced by local pixels only. Therefore, it is hard for 2D Gaussians to dynamically reallocate to regions requiring finer-grained details. The Gaussians would be trapped in local minima, leading to redundant Gaussians that are optimized as either low opacity or excessively large size, resulting in artifacts in rendering results. 

In standard 3D Gaussian Splatting pipeline \cite{kerbl3Dgaussians}, Gaussians with low opacity or excessive size are pruned, whereas those with high gradient responses are split into two. However, this strategy is not directly applicable to Bézier curves. Note that in volumetric representations, large Gaussians are usually unnecessary for modeling any particular structure. In contrast, VG representations often encompass large uniform regions such as backgrounds or areas with homogeneous colors (\textit{e.g.,} walls). Consequently, the size-based pruning strategy would remove critical structures in VG representations. Similarly, splitting a Bézier curve into two can introduce significant randomness, as high-gradient regions do not always indicate poor reconstruction in VGs. Since VGs assume a uniform color within each enclosed region, complex textures naturally produce high gradients. This does not imply the region should be split, as it may disrupt the semantic consistency of the VG representation.

To address this issue, this work introduces a new pruning and densification strategy to dynamically adjust the density of Bézier curves throughout the optimization process. For pruning redundant Bézier curves, we apply three criteria to ensure a stable and precise optimization process. First, we remove curves with opacity below a dynamic threshold that gradually decreases as optimization progresses, ensuring that weakly contributing curves are eliminated while preserving essential structures. Second, we remove curves with an area below a predefined threshold, as they contribute minimally to the final representation. To remove visually insignificant or noisy Bézier curves, we apply an opacity-based filtering strategy: for per-segment opacities, we discard curves whose middle segment is significantly fainter than both ends—often indicating that a single curve improperly spans a region better represented by multiple curves; otherwise, we discard curves with overall opacity below a threshold (e.g., 0.2) to eliminate globally low-visibility curves. Third, we remove curves that exhibit high color similarity with surrounding curves and have significant overlap, as they provide little additional information and can be pruned without affecting the final representation. By removing such curves, we allow the optimization to introduce more appropriate replacements, leading to a more accurate and visually coherent vectorized representation. For curve densification, we adopt an error-driven curve allocation strategy inspired by LIVE \cite{xu2022live}. Specifically, we compute connected error regions, rank them by area, and add new Bézier curves into the highest-error regions. This adaptive redistribution mechanism ensures that curves are allocated to where they would contribute the most to the reconstruction fidelity. The pruning and densification strategy enables a ``global receptive field'' for redistributing curve density, effectively preserving visual details for high-fidelity rendering while maintaining a compact vector representation.
% \begin{table}[t]
% \centering
% \caption{Forward and backward speed of DiffVG \cite{Li:2020:DVG} and Bézier splatting (tested on a 2,040$\times$1,344 image, using 2,048 curves).}
% \vspace{-1em}
% \resizebox{0.95\linewidth}{!}{%
% \begin{tabular}{ll|ccc}
% \hline
%  & & DiffVG & Bézier splatting & Speedup \\
% \hline
% \multirow{2}{*}{Open} & Forward & 141.3ms  & \textbf{7.2ms} & 19.6$\times$ \\  
% & Backward & 701.3.ms & \textbf{4.7ms} & 149.2$\times$ \\  
% \hline
% \multirow{2}{*}{Closed} & Forward & 85.2ms  & \textbf{14.2ms} & 6.0$\times$ \\  
% & Backward & 448.3ms  & \textbf{15.7ms} & 28.5$\times$ \\  
% \hline
% \end{tabular}
% }
% \label{tab:speed_comparison}
% \vspace{-.5em}
% \end{table}

\begin{table}[t]
\centering
\caption{Computational speed for rendering a 2,040$\times$1,344 image with 2,048 curves.}
\vspace{-.7em}
\resizebox{0.83\linewidth}{!}{%
\begin{tabular}{l|ccc|ccc}
\hline
& \multicolumn{3}{c|}{\textbf{Open curve}} & \multicolumn{3}{c}{\textbf{Closed curve}} \\
& DiffVG \cite{Li:2020:DVG} & Bézier Splatting & Speedup & DiffVG \cite{Li:2020:DVG} & Bézier Splatting & Speedup \\
\hline
Forward pass & 141.3ms & \textbf{4.5ms} & \textit{31.4}$\times$ & 85.2ms & \textbf{14.1ms} & \textit{6.0}$\times$ \\
Backward pass & 701.3ms & \textbf{4.7ms} & \textit{149.2}$\times$ & 448.3ms & \textbf{24.58ms} & \textit{18.2}$\times$ \\
\hline
\end{tabular}
}
\label{tab:speed_comparison}
% \vspace{-.5em}
\end{table}

\begin{figure}[!t]
\centering
% \vspace{-2em}
\includegraphics[width=0.98\linewidth]{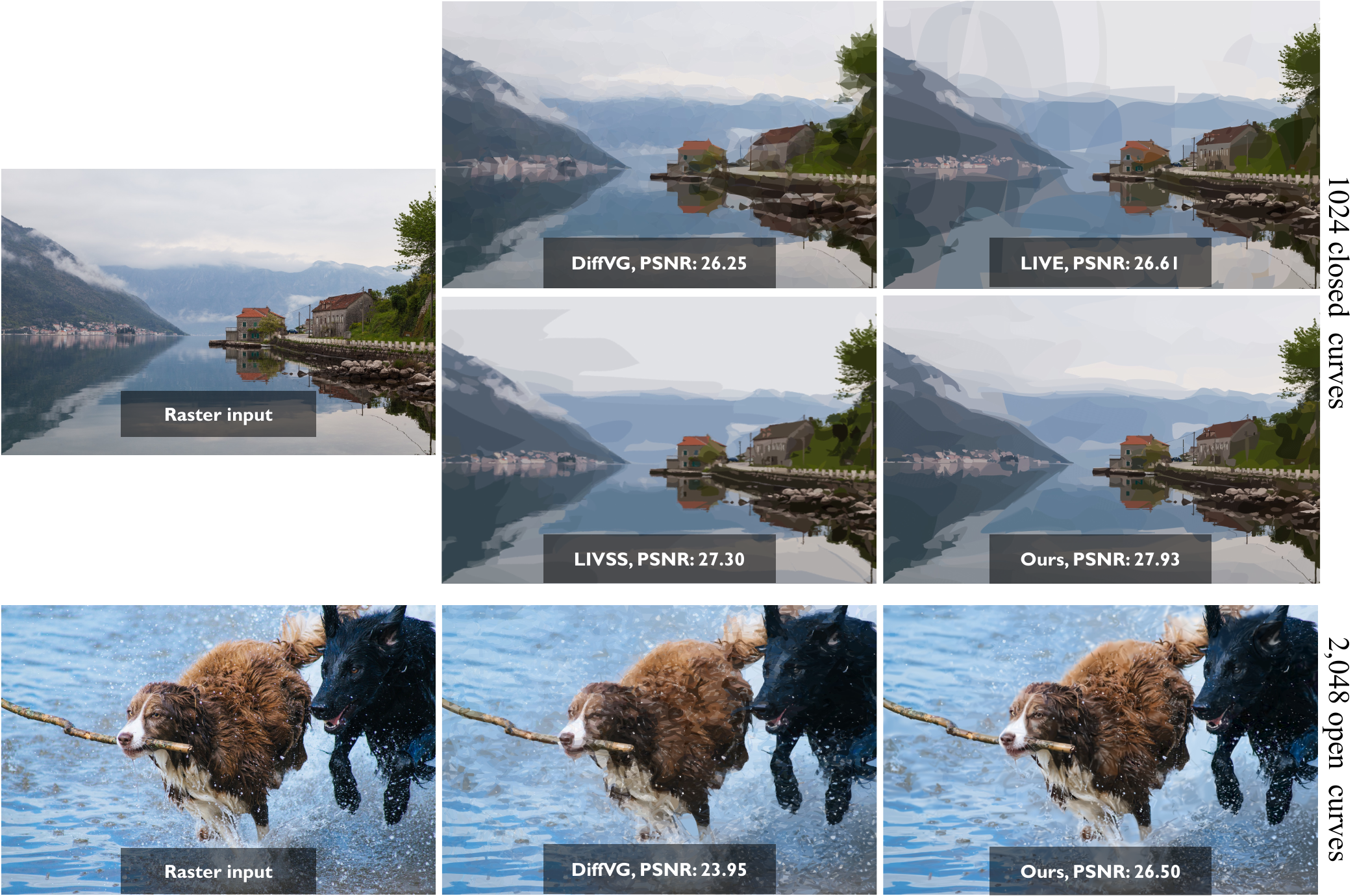}
\vspace{-.5em}
\caption{A qualitative comparison of our method and the state-of-the-art differentiable VG rasterization methods, including DiffVG \cite{Li:2020:DVG},  LIVE \cite{xu2022live}, and LIVSS \cite{livss}.
% , with different number ($N$) of open and closed curves. 
% The raster input is taken from the DIV2K dataset \cite{Timofte_2017_CVPR_Workshops}.
}
\label{fig:comparison}
\vspace{-.5em}
\end{figure}

\vspace{-.5em}
\section{Experiments}
\vspace{-.5em}

\subsection{Experimental Setups}
\noindent\textbf{Implementation details.} We implement Bézier Splatting in PyTorch \cite{paszke2019pytorch} and optimize it by using the Adam optimizer \cite{2015-kingma} with a StepLR learning rate scheduler. The learning rate is initialized at 0.01 for color, 2e-4 for Bézier curve control points, and 0.1 for opacity. For the pruning and densification strategy, the opacity threshold is set to 0.02, and the overlap threshold based on Axis-Aligned Bounding Boxes (AABB) is set to 0.9. Following the approach in LIVE \cite{xu2022live}, new curves are initialized in a circular pattern, and the number of added curves matches the number of removed ones to maintain a constant total curve count. Open curves are optimized for 15,000 iterations, and closed curves are optimized for 10,000 iterations. Pruning and densification are applied every 400 steps until the last but 1,000 steps, after which they are halted to stabilize the representation. 

\begin{figure*}[t]
\vspace{-.5em}
\centering
\includegraphics[width=1\linewidth]{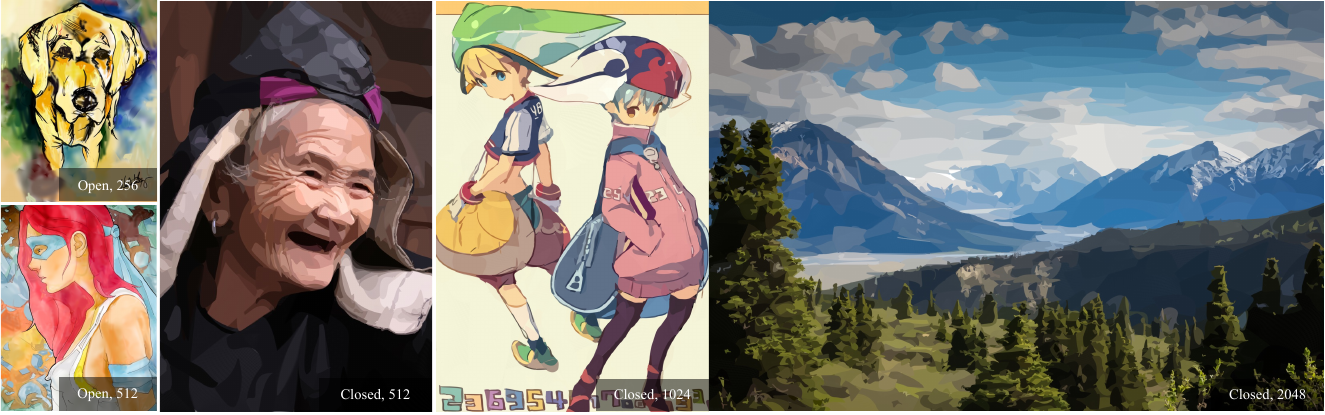}
\vspace{-1.5em}
\caption{
Our Bézier Splatting achieves high-quality image vectorization results for various types of images including artworks, cartoons, and natural images. Curve type and count are indicated at the bottom right of each sample.
% \caption{We showcase the image vectorization results by our Bézier Splatting, with open or closed curves and different curve counts indicated at the bottom right of each example. The raster inputs, shown at the top left of each example, are taken from the DIV2K \cite{Timofte_2017_CVPR_Workshops}, KIOAK \cite{kodak1999}, Clipart \cite{inoue2018cross}, and Danbooregions \cite{DanbooRegion2020} datasets. Zoom in for a better view of details.
}
\label{fig:showcase}
% \vspace{-1em}
\end{figure*}

\noindent\textbf{Datasets.} We comprehensively evaluate our method across different image domains. We use the publicly available DIV2K \cite{Timofte_2017_CVPR_Workshops} dataset for evaluating natural images. Due to the high computational cost of baseline method LIVE \cite{xu2022live}, we uniformly subsample the DIV2K dataset by selecting one out of every four images, resulting in a final evaluation set of 200 images from the original 800-image dataset.
Additionally, we test our method on non-photorealistic images, including the artwork images from Clipart1K dataset \cite{inoue2018cross} and cartoon images from Danbooregions dataset \cite{DanbooRegion2020}, to demonstrate its effectiveness for diverse types of images, as shown in Fig. \ref{fig:showcase}. 

\begin{table*}[!t]
\centering
\caption{
    \label{tab:comparisons_div2k} Quantitative evaluation and optimization efficiency of differentiable VG methods on the DIV2K dataset \cite{Timofte_2017_CVPR_Workshops}.
}
\setlength{\tabcolsep}{3pt}
\resizebox{0.99\textwidth}{!}{
\begin{tabular}{l|l|cccc|cccc|cccc}
\hline
& \multirow{2}{*}{\textbf{Method}} 
& \multicolumn{4}{c|}{\textbf{256 curves}} 
& \multicolumn{4}{c|}{\textbf{512 curves}} 
& \multicolumn{4}{c}{\textbf{1024 curves}} \\
& & $SSIM^\uparrow$ & $PSNR^\uparrow$ & $LPIPS^\downarrow$ & \cellcolor{timegray}$\text{Opt.}^\downarrow$
  & $SSIM^\uparrow$ & $PSNR^\uparrow$ & $LPIPS^\downarrow$ & \cellcolor{timegray}$\text{Opt.}^\downarrow$
  & $SSIM^\uparrow$ & $PSNR^\uparrow$ & $LPIPS^\downarrow$ & \cellcolor{timegray}$\text{Opt.}^\downarrow$ \\
\hline

\multirow{2}{*}{Open} 
& DiffVG~\cite{Li:2020:DVG} 
  & \cellcolor{baselinegray}0.552 & \cellcolor{baselinegray}19.83 & \cellcolor{baselinegray}0.563 & \cellcolor{timegray}18.9min 
  & \cellcolor{baselinegray}0.587 & \cellcolor{baselinegray}21.47 & \cellcolor{baselinegray}0.537 & \cellcolor{timegray}22.0min 
  & \cellcolor{baselinegray}0.616 & \cellcolor{baselinegray}22.62 & \cellcolor{baselinegray}0.517 & \cellcolor{timegray}30.6min \\
& Ours 
  & \cellcolor{oursblue}\textbf{0.600} & \cellcolor{oursblue}\textbf{22.17} & \cellcolor{oursblue}\textbf{0.540} & \cellcolor{timegray}\textbf{3.4min}
  & \cellcolor{oursblue}\textbf{0.646} & \cellcolor{oursblue}\textbf{23.79} & \cellcolor{oursblue}\textbf{0.498} & \cellcolor{timegray}\textbf{3.3min}
  & \cellcolor{oursblue}\textbf{0.699} & \cellcolor{oursblue}\textbf{25.45} & \cellcolor{oursblue}\textbf{0.448} & \cellcolor{timegray}\textbf{3.2min} \\
\hline

\multirow{4}{*}{Closed} 
& DiffVG~\cite{Li:2020:DVG}
  & \cellcolor{baselinegray}0.578 & \cellcolor{baselinegray}20.69 & \cellcolor{baselinegray}0.548 & \cellcolor{timegray}16.4min
  & \cellcolor{baselinegray}0.601 & \cellcolor{baselinegray}21.82 & \cellcolor{baselinegray}0.531 & \cellcolor{timegray}18.5min
  & \cellcolor{baselinegray}0.631 & \cellcolor{baselinegray}22.95 & \cellcolor{baselinegray}0.509 & \cellcolor{timegray}25.1min \\
& LIVE~\cite{xu2022live} 
  & \cellcolor{baselinegray}0.576 & \cellcolor{baselinegray}20.09 & \cellcolor{baselinegray}0.543 & \cellcolor{timegray}2.6h
  & \cellcolor{baselinegray}0.611 & \cellcolor{baselinegray}21.70 & \cellcolor{baselinegray}\textbf{0.521} & \cellcolor{timegray}4.2h
  & \cellcolor{baselinegray}0.648 & \cellcolor{baselinegray}23.11 & \cellcolor{baselinegray}0.495 & \cellcolor{timegray}5.1h \\
& LIVSS~\cite{livss} 
  & \cellcolor{baselinegray}\textbf{0.586} & \cellcolor{baselinegray}17.71 & \cellcolor{baselinegray}\textbf{0.542} & \cellcolor{timegray}39.2min
  & \cellcolor{baselinegray}\textbf{0.630} & \cellcolor{baselinegray}18.71 & \cellcolor{baselinegray}0.530 & \cellcolor{timegray}54.3min
  & \cellcolor{baselinegray}\textbf{0.678} & \cellcolor{baselinegray}19.83 & \cellcolor{baselinegray}0.517 & \cellcolor{timegray}1.4h \\
& Ours 
  & \cellcolor{oursblue}0.580 & \cellcolor{oursblue}\textbf{20.74} & \cellcolor{oursblue}0.546 & \cellcolor{timegray}\textbf{7.8min}
  & \cellcolor{oursblue}0.607 & \cellcolor{oursblue}\textbf{22.11} & \cellcolor{oursblue}0.528 & \cellcolor{timegray}\textbf{8.3min}
  & \cellcolor{oursblue}0.639 & \cellcolor{oursblue}\textbf{23.45} & \cellcolor{oursblue}\textbf{0.507} & \cellcolor{timegray}\textbf{8.6min} \\
\hline
\end{tabular}
}
\end{table*}

\vspace{-.5em}
\subsection{Method Comparison}
\vspace{-.5em}

Table \ref{tab:speed_comparison} reports the forward and backward runtime for processing 2,048 curves on an image with a resolution of 2,040$\times$1,344. Compared to DiffVG \cite{Li:2020:DVG}, our method significantly accelerates VG rasterization by 31.4$\times$ faster per forward step and 149.2$\times$ faster per backward step for open curves. For closed curves, our color filling strategy requires sampling 20 additional Bézier curves per closed curve, while our method remains highly efficient, achieving 6$\times$ faster forward and 18.2$\times$ faster backward computation.

Table \ref{tab:comparisons_div2k} quantitatively evaluates the quality of differentiable VG representations by three commonly used metrics, MS-SSIM \cite{583d8b59c51a4f3e9bd0b90731111848}, PSNR, and LPIPS \cite{zhang2018perceptual}. Our method demonstrates higher optimization efficiency and rendering fidelity for both open and closed curves with different curve numbers.

Fig. \ref{fig:comparison} visually compares DiffVG \cite{Li:2020:DVG}, LIVE \cite{Du:2023:IVE}, \cite{livss}, and our Bézier Splatting under 512 closed curves and 2048 open curves. 
% As the number of paths increases, it shows progressive enhancement in the visual details. 
Compared to DiffVG \cite{Li:2020:DVG}, our method captures significantly more fine-grained textures. Compared to LIVE \cite{xu2022live}, our method effectively enhances rendering quality, resulting in higher-fidelity results. This improvement stems from our method's ability to globally optimize all curves, rather than LIVE's layer-by-layer path-adding strategy. Compared to LIVSS \cite{livss}, our method demonstrates superior ability in preserving fine structural details, especially in text regions. Due to its heavy reliance on layer-wise semantic simplification, LIVSS struggles to optimize regions where semantic information is ambiguous or difficult to extract. Moreover, by optimizing the entire VG simultaneously, our approach ensures a cohesive and natural appearance, avoiding the accumulation of errors and inconsistencies introduced by layer-wise updates.

As shown in Fig. \ref{fig:showcase}, Bézier Splatting consistently renders high-quality VGs across various image domains from photorealistic natural images, watercolor paintings, to cartoon images. Unlike learning-based image vectorization methods \cite{lopes2019learned, reddy2021im2vec, Cao_2023_CVPR}, which can not easily generalize to out-of-domain data, our method can flexibly handle different types of images. 

\begin{figure*}[!t]
\centering
% \vspace{-2em}
\includegraphics[width=0.92\linewidth]{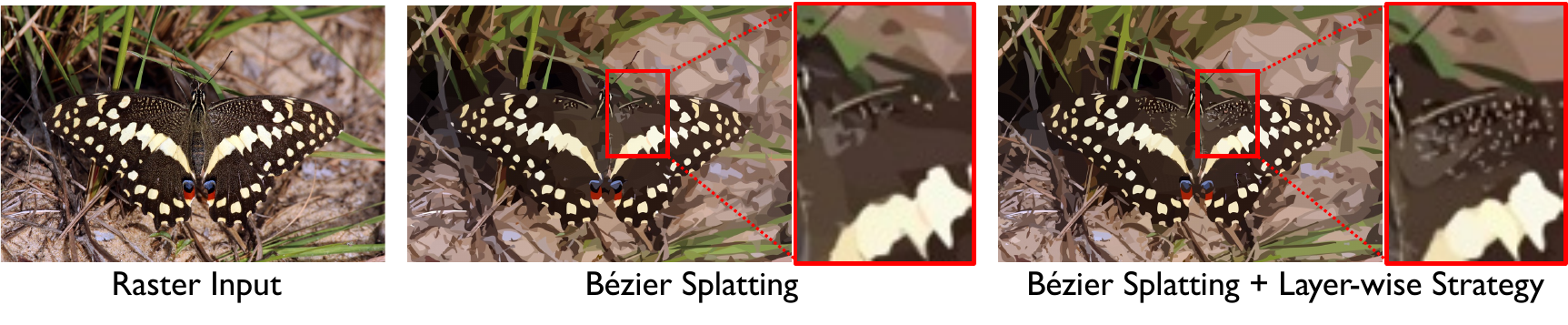}
\vspace{-1em}
\caption{As a differentiable VGs renderer, Bézier Splatting can integrate with topology-aware strategies such as layer-wise vectorization \cite{xu2022live} to further improve compositionality and detail.}
\label{fig:layer}
% \vspace{-.5em}
\end{figure*}

\begin{figure*}[!t]
\centering
% \vspace{-2em}
\includegraphics[width=1\linewidth]{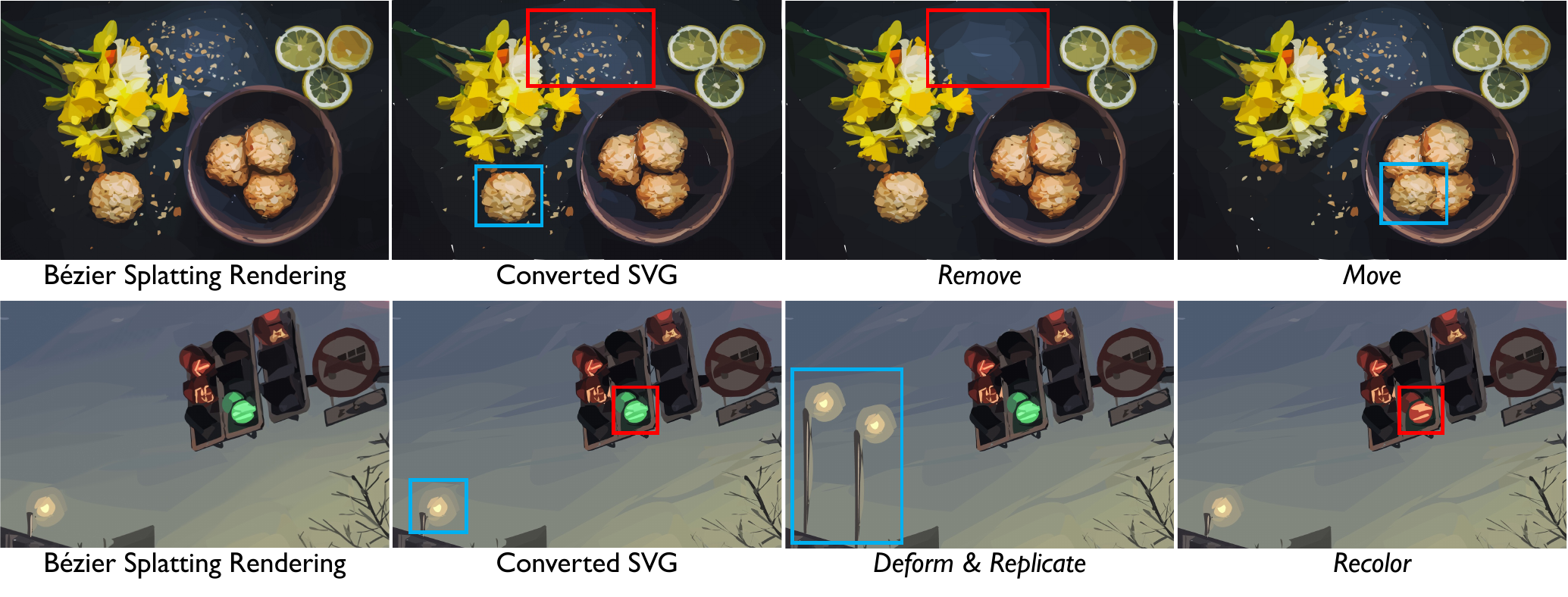}
\vspace{-2em}
\caption{Our Bézier Splatting representation is compatible with the standard SVG XML format and supports flexible vector editing operations.}
\label{fig:editing}
\vspace{-.5em}
\end{figure*}

\vspace{-.5em}
\subsection{Layer-wise Image Vectorization}
Existing works such as LIVE \cite{xu2022live}, SGLIVE \cite{zhou2024segmentation} and LIVSS \cite{livss} have demonstrated that layer-wise vectorization strategies can significantly improve the topology and compositionality of DiffVG-generated vector graphics \cite{Li:2020:DVG}.
Since the proposed Bézier Splatting is a fast and differentiable VGs renderer alternative to DiffVG, is naturally compatible with these topology-aware strategies and can leverage them to further enhance the structural quality of the optimized vector outputs.
As shown in Fig. \ref{fig:layer}, we incorporate the layer-wise vectorization approach from LIVE \cite{xu2022live} into the Bézier Splatting optimization process
This integration yields improved compositional integrity and finer details for regions with complex patterns such as the butterfly's wings.

\vspace{-.5em}
\subsection{Editability and Compatibility with SVG XML Format}
Our Bézier Splatting representation is fully compatible with the standard SVG XML format, ensuring the same editability as standard SVGs and seamless integration with existing SVG tools.
As shown in Fig. \ref{fig:editing}, a high-fidelity SVG can be obtained via a simple conversion algorithm (see Appendix for details).
Once converted to SVG, the vectorized output is fully editable at the primitive level, allowing users to freely deform, remove, move, replicate, and recolor individual elements.

% \subsection{Ablation study}
% Please refer to Appendix. \ref{sec:ablation} for the ablation study.

\vspace{-.5em}
\section{Conclusion}
\vspace{-.5em}

This work has presented Bézier Splatting, a novel differentiable vector graphics (VGs) representation that leverages Gaussian splatting for efficient Bézier curve optimization. Our method achieves 30$\times$ faster forward computation and 150$\times$ faster backward computation in rasterization compared to existing methods, while also delivering high rendering fidelity. Additionally, our adaptive pruning and densification strategy improves optimization by dynamically adjusting curve placement during optimization. Extensive experiments have demonstrated that Bézier Splatting outperforms existing differentiable VG methods in both training efficiency and visual quality, making it a promising solution for scalable applications of VGs. 

\noindent\textbf{Limitation and future work.}
The closed curves in Bézier Splatting are enforced to have convex shapes using Xing loss \cite{xu2022live} to prevent false interpolation results, which may slightly reduce the capacity of VG representations. Additionally, the closed curves require sampling more Gaussian points to represent area boundaries precisely, leading to slower computation compared to open curves.
An interesting future direction is to leverage the efficiency and differentiability of our approach for high-quality VG synthesis applications such as text-to-VG generation or text-guided VG editing.

\section{Acknowledgement}
This work was supported in part by the National Science Foundation (NSF) and SC EPSCoR Program under award number 
\#OIA-2242812.
This research used in part resources on the Palmetto Cluster at Clemson University under NSF awards MRI 1228312, II NEW 1405767, MRI 1725573, and MRI 2018069. The views expressed in this article do not necessarily represent the views of NSF or the United States government. The authors also thank Dr. Yiqi Zhong and Dr. Luming Liang  for beneficial discussions.

% \section*{References}

{
    % \small
    \bibliographystyle{ieeenat_fullname}
    \bibliography{main}
}

%%%%%%%%%%%%%%%%%%%%%%%%%%%%%%%%%%%%%%%%%%%%%%%%%%%%%%%%%%%%
\newpage
\appendix

\appendix
% 1. describe line is multi opacities, curves is single opacity.@xi Liu
% 2.1 ablation study: ori vs  adc vs live (SSIM, LPIPS, PSNR, training time)
% 2.2 loss ablation (delete loss sperately)
% 2.3 forward and backward time for different area curves resolution, and time statics of each forward module. backward time for different loss.
% 3.1 More examples for different paths @ chaoyi
% 3.3 More examples for different datasets (optional quantity) @ Xi and choose the demo for appendix @chaoyi 
% 3.3 training process of our LIVE
% 3.4 Shape analysis @Chaoyi
% 3.5 Our + LIVSS for DIV2K @Chaoyi
% 3.6 Convert our to XML SVG @Chaoyi

% 4. website
% 4.1 Changing teaser
% 4.2 Changing gallery
% 4.3 Updating comparison
% 4.4 Video of Training process Live and our
% 4.5 Updating Efficiency comparison
% 4.6 SVG link (Optional: Editting Based on LLIVS)

\section{Ablation Study}
% \begin{figure*}[!t]
% \centering
% \includegraphics[width=0.8\linewidth]{figures/ablations.pdf}
% \vspace{-1em}
% \caption{Ablation study on the proposed adaptive pruning and densification strategy and multi-opacity strategy. 
% }
% \vspace{-.5em}
% \label{fig:ablation}
% \end{figure*}

\label{sec:ablation}

\begin{figure*}[h]
\centering
\includegraphics[width=0.9\linewidth]{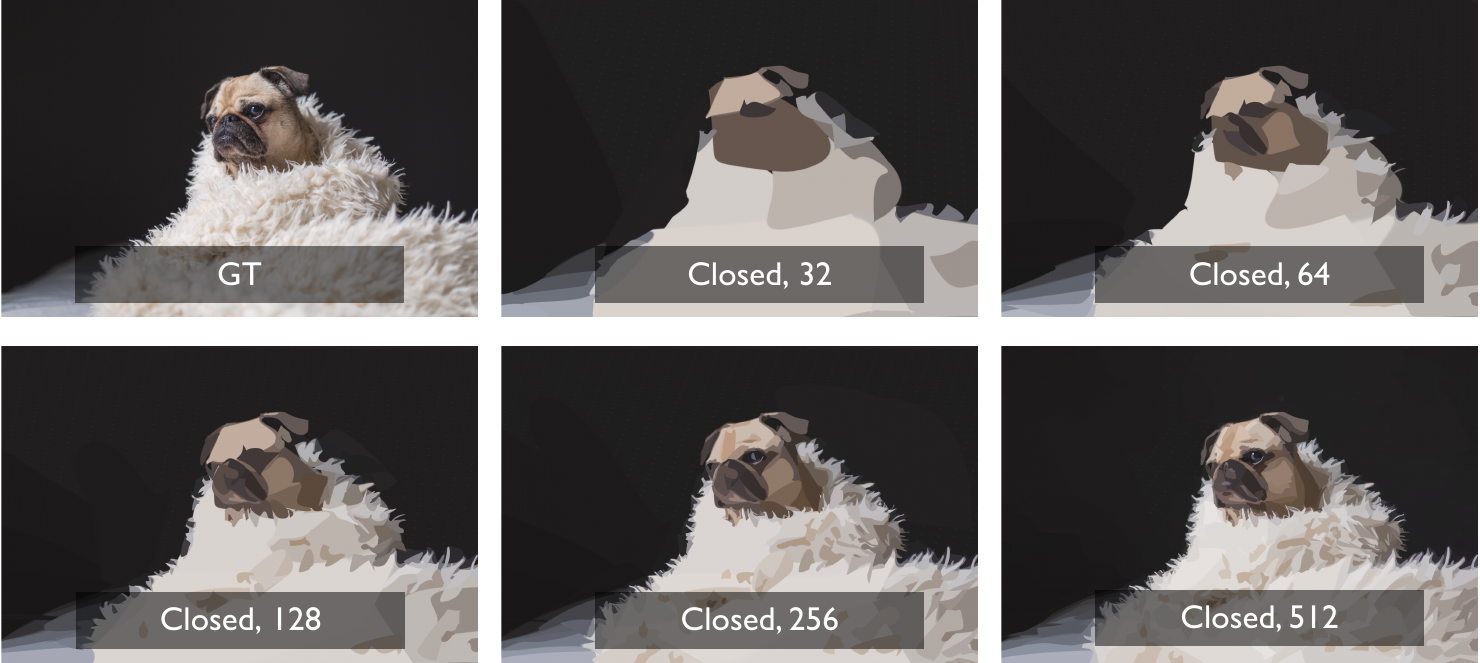}
% \vspace{-2em}
\caption{Our Bézier Splatting is fully compatible with the layer-wise vectorization strategies \cite{xu2022live}. }
\label{fig:layerwise-training}
\end{figure*}

\noindent\textbf{The effectiveness of adaptive pruning and densification and layer-wise vectorization strategy.}
We conduct an ablation study to evaluate the effectiveness of our adaptive pruning and densification strategy. Quantitative results are shown in Table~\ref{tab:adc_ablation_table}. To further investigate the flexibility of our framework, we incorporate a layer-wise curve addition strategy inspired by LIVE~\cite{xu2022live}, where curves are progressively added during training. Although this strategy achieves slightly better reconstruction metrics, it doubles the training time compared to our adaptive pruning and densification approach, making it less favorable for time-sensitive applications. Fig. \ref{fig:layerwise-training} shows our result with layer-wise image vectorization strategy.

\definecolor{bestrow}{RGB}{235,245,255}

\begin{table}[!h]
\centering
\caption{An ablation study on adaptive pruning and densification and layer-wise training strategy, evaluated on 512 closed curves with DIV2K dataset \cite{Timofte_2017_CVPR_Workshops}.}
% \vspace{-.5em}
\resizebox{0.7\linewidth}{!}{%
\begin{tabular}{l|cccc}
\hline
\textbf{Method} & $SSIM^\uparrow$ & $PSNR^\uparrow$ & $LPIPS^\downarrow$ & $Opt^\downarrow$ \\
\hline
No Strategy & 0.590 & 21.10 & 0.530 & \textbf{8.3 min} \\
% \rowcolor{bestrow}
~~-- w/ Prune\&Densify & 0.607 & 22.11 & 0.528 & 8.3 min \\
~~-- w/ Layer-wise & \textbf{0.613} & \textbf{22.21} & \textbf{0.521} & 16.2 min \\
\hline
\end{tabular}
}
\label{tab:adc_ablation_table}
\end{table}

\begin{figure}[h]
\centering
\includegraphics[width=0.9\linewidth]{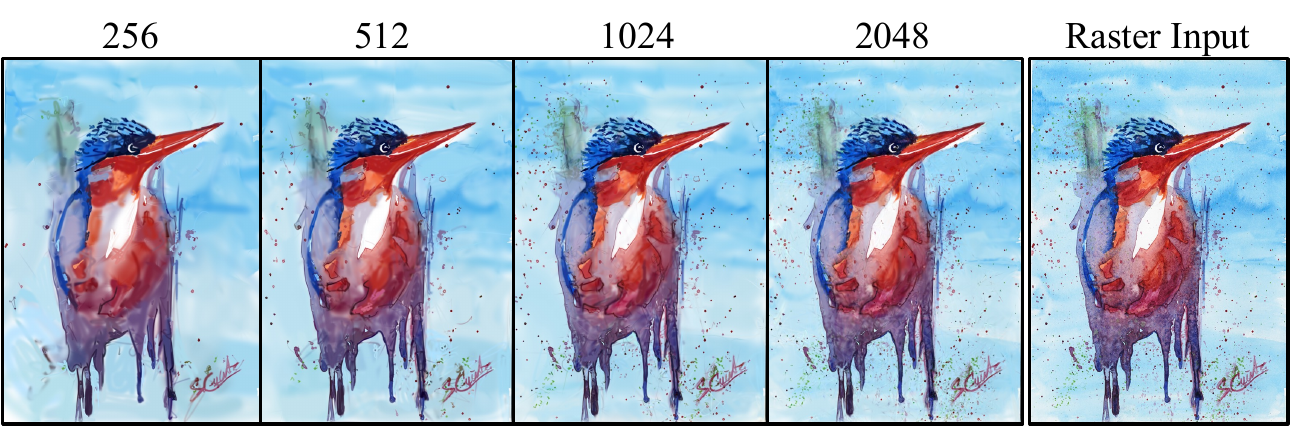}
\vspace{-.5em}
\caption{A comparison of different curve numbers by Bézier Splatting. 
}
\label{fig:curve number}
\end{figure}

\noindent\textbf{Comparing different numbers of curves.} We progressively increase the number of curves for vectorizing a watercolor image (Fig. \ref{fig:curve number}) with numerous small spots. As the number of curves grows, our approach first reconstructs the foreground object, the bird, then gradually refines the smaller spots. This demonstrates that our method prioritizes the primary structure before optimizing finer details, effectively distributing curves to balance global structure and local texture representation, thanks to our adaptive pruning and densification strategy.

\noindent\textbf{A systematic evaluation of computation time.} To systematically evaluate the computational efficiency of our framework, we report detailed timing statistics across varying configurations of interpolation curve numbers and curve resolutions, as summarized in Table~\ref{tab:forward_backward_decomposed}. The total forward time includes two major parts: the Gaussian sampling and the Gaussian splatting. Increasing the number of interpolated curves within the closed Bézier curves leads to a near-linear growth in both forward and backward runtimes, demonstrating good computational scalability. When fixing the interpolation number (\textit{e.g.}, at 40) and varying the number of Bézier curves, the forward time increases moderately, reflecting the additional cost of handling finer spatial detail. The splatting stage dominates the forward time at higher number of curves due to the increased number of sampled points per region, while the Gaussian sampling time is the primary bottleneck at lower number of curves. This is attributed to overheads from sequential memory allocation and data transfer operations, which are not effectively parallelized on the GPU, suggesting potential for further optimization.

\begin{table}[!t]
\centering
\caption{A systematic evaluation of computation time. We test the forward and backward time (ms) under varying numbers of sampled curves within one closed Bézier curve (Inter. \#), and the total number of Bézier curves (Curve \#).}
\resizebox{1.0\linewidth}{!}{%
\begin{tabular}{ccccccc}
\toprule
\textbf{Inter. \#} & \textbf{Curve \#} & \textbf{Forward w/ gradient} & \textbf{Sample Gaussians} & \textbf{Splatting} & \textbf{Backward} & \textbf{Render (FPS)} \\
\midrule
20   & 2048 & 10.66 & 4.12   & 6.11   & 21.64 & 103.45 \\
40   & 2048 & 14.79 & 5.11   & 9.32   & 25.06 & 68.30  \\
80   & 2048 & 27.01 & 8.22   & 17.74  & 33.40 & 37.90  \\
\midrule
40   & 256  & 7.20  & 3.81 & 3.09  & 16.72 & 150.30 \\
40   & 512  & 7.70  & 3.88 & 3.67 & 17.42 & 144.30 \\
40   & 1024 & 10.39 & 4.06  & 5.96 & 19.96 & 114.30 \\
40   & 2048 & 14.79 & 5.11   & 9.32   & 25.06 & 68.30  \\
\bottomrule
\end{tabular}
}
\label{tab:forward_backward_decomposed}
\end{table}

\begin{table}[h]
\centering
\caption{A quantitative comparison between Adobe Image Trace \cite{illustrator2025} and our method across different images.}
\resizebox{1.0\linewidth}{!}{%
\begin{tabular}{c|cc|cc|cc|cc|cc}
\toprule
\textbf{Image ID} & \multicolumn{2}{c|}{0004} & \multicolumn{2}{c|}{0008} & \multicolumn{2}{c|}{0012} & \multicolumn{2}{c|}{0016} & \multicolumn{2}{c}{0020} \\
\midrule
\textbf{Method} & Adobe \cite{illustrator2025} & Ours & Adobe \cite{illustrator2025}& Ours & Adobe \cite{illustrator2025}& Ours & Adobe \cite{illustrator2025}& Ours & Adobe \cite{illustrator2025}& Ours \\
\midrule
\textbf{$SSIM^\uparrow$} & 0.835 & 0.849 & 0.645 & 0.637 & 0.658 & 0.658 & 0.627 & 0.628 & 0.814 & 0.822 \\
\textbf{$PSNR^\uparrow$} & 26.43 & 28.62 & 23.26 & 24.33 & 22.29 & 22.87 & 23.87 & 24.54 & 26.67 & 28.31 \\
\textbf{$LPIPS^\uparrow$} & 0.380 & 0.387 & 0.489 & 0.489 & 0.523 & 0.518 & 0.534 & 0.543 & 0.505 & 0.503 \\
\bottomrule
\end{tabular}
}
\label{tab:traditional_comparsion}
\end{table}

\noindent\textbf{Comparison with conventional image vectorization methods.}  We conduct a comparison against the Image Trace from Adobe Illustrator \cite{illustrator2025} on five images from the DIV2K dataset. Both methods are evaluated using the same number of parameters (around 20K). As shown in Table~\ref{tab:traditional_comparsion}, our method achieves higher fidelity, with an average PSNR of 25.734 db compared to 24.904 db of Image Trace.

\noindent\textbf{Arbitrary resolution rendering.}  
Our Bézier Splatting representation naturally supports rendering at arbitrary resolutions because all Gaussian parameters are analytically derived from the underlying Bézier curves. When higher-resolution images are required, proportionally increasing the sampling rate is sufficient to maintain reconstruction fidelity. For example, doubling the resolution simply doubles the sampling density. As shown in Table~\ref{tab:resolution}, rendering at higher resolutions (4K or 8K) with increased sampling preserves a comparable level of visual quality to the 2K baseline, without introducing noticeable artifacts. We report results on the first four images from the \cite{Timofte_2017_CVPR_Workshops} to illustrate this property.

\begin{table}[h]
\centering
\caption{
Results of Bézier Splatting rendered at different resolutions and sampling settings.  
Notation: ``orig.'' denotes the baseline sampling rate used during optimization, 
and ``$\times$S'' indicates S-times higher sampling density.
}
\label{tab:resolution}
% \vspace{0.3em}
\resizebox{0.6\linewidth}{!}{%
\begin{tabular}{lcccccc}
\toprule
\multirow{2}{*}{\textbf{Image ID}} & 
\multirow{2}{*}{\textbf{2K (orig.)}} &
\multicolumn{2}{c}{\textbf{4K}} & 
\multicolumn{2}{c}{\textbf{8K}} \\
\cmidrule(lr){3-4} \cmidrule(lr){5-6}
 &  & \textbf{orig.} & \textbf{($\times$2S)} & \textbf{orig.} & \textbf{($\times$4S)} \\
\midrule
0004 & 26.8976 & 27.1553 & 27.2761 & 25.9800 & 26.6594 \\
0008 & 21.2063 & 21.0614 & 21.4340 & 19.8318 & 21.1214 \\
0012 & 19.8816 & 19.9079 & 20.0900 & 18.9389 & 19.7760 \\
0016 & 22.3456 & 22.4327 & 22.5408 & 21.7088 & 22.3140 \\
\bottomrule
\end{tabular}
}
\end{table}

% \begin{table}[h]
% \centering
% \caption{Forward and backward time (ms) under different number of interplation curves  for 2048 closed curves.}
% \begin{tabular}{lccc}
% \toprule
% \textbf{Curve num} & \textbf{Forward Time (ms)} & \textbf{Backward Time (ms)} & \textbf{FPS} \\
% \midrule
% 20    & 10.66 (4.1, 6.1)                      & 21.64   &103.45                      \\
% 40  &14.79 (5.1, 9.3)                      & 25.06  & 68.3                         \\
% 80  & 27.01 (8.2, 17.7)                      & 33.4 & 37.9                     \\
% \bottomrule
% \end{tabular}
% \label{tab:forward_backward_resolution}
% \end{table}

% \begin{table}[h]
% \centering
% \caption{Forward and backward time (ms) under different curve resolutions with same interplation curves number: 40.}
% \begin{tabular}{lccc}
% \toprule
% \textbf{Curve Resolution} & \textbf{Forward Time (ms)} & \textbf{Backward Time (ms)} & \textbf{FPS}\\
% \midrule
% 256 curves    &7.2 (3.807, 3.09)                       & 16.724        & 150.3       \\
% 512 curves   &7.7 (3.877, 3.669)                       & 17.424        & 144.3        \\
% 1024 curves  &10.39 (4.06, 5.958)                       & 19.96        & 114.3        \\
% 2048 curves  &14.79 (5.1, 9.3)                      & 25.06  & 68.3                         \\
% \bottomrule
% \end{tabular}
% \label{tab:forward_backward_resolution}
% \end{table}

\section{Details of Adaptive Pruning and Densification}
To provide a clearer understanding of our optimization behavior, we present here the detailed procedure of the pruning–densification algorithm \ref{alg:adaptive_pruning_densification} used in our Bézier-splatting framework.This algorithm is designed to mitigate the local-minima issue observed in 3D Gaussian Splatting (3DGS) and is adapted in our Bézier-splatting framework, where suboptimal primitive initialization leaves certain regions insufficiently covered.
Inspired by the pruning–splitting mechanism of 3DGS [1], our method dynamically reallocates Bézier curves according to the reconstruction error map: curves with low opacity, small area, or high overlap with nearby curves of similar color are pruned, while new curves are inserted into regions with high reconstruction error to improve coverage.
This reallocation keeps the total curve budget fixed, enhances reconstruction quality, and helps the optimization escape poor local minima.
For open curves, we additionally introduce a splitting rule:
if the middle opacity of a curve is more than 0.5 lower than that of both endpoints, the curve is split into two segments to better adapt to local structure variations.

\begin{algorithm}[t]
\caption{\textbf{Adaptive Pruning and Densification for Closed Bézier Curves}}
\label{alg:adaptive_pruning_densification}
\begin{algorithmic} % no [1] => no line numbers

% ---- line between caption and Notation ----

% ---- Notation (operators in bold), one per line ----
\STATE \textbf{Notation:}
\STATE \quad $\mathbf{opacity}(b)$ — mean opacity of curve $b$.
\STATE \quad $\mathbf{area}(b)$ — area enclosed by curve $b$.
\STATE \quad $\mathbf{colordiff}(b_i,b_j)$ — Euclidean color difference between curves $b_i$ and $b_j$.
\STATE \quad $\mathbf{IoU}(b_i)$ — intersection-over-union between the bounding box of $b_i$ and neighboring curves with $\mathbf{colordiff}<0.03$.
\STATE \quad $\mathbf{ConnectedComponents}(E)$ — connected error regions extracted from the quantized error map.

% ---- line below Notation ----
\STATE {\rule{\linewidth}{0.5pt}}

% ---- Inputs / Outputs ----
\STATE \textbf{Input}~: Closed Bézier curve set $\mathcal{B}=\{b_1,\dots,b_N\}$; opacity map $\alpha$; error map $E$; iteration index $t$.
\STATE \textbf{Output}: Updated curve set $\mathcal{B}'$.

% ---- Algorithm body ----
\STATE Initialize $\mathcal{B}' \leftarrow \mathcal{B}$.

\STATE \textbf{Pruning phase}
\FOR{$i \gets 1$ \TO $N$}
  \IF{$\mathbf{opacity}(b_i) < \tau_{\mathbf{opacity}}(t)$}
    \STATE Remove $b_i$ from $\mathcal{B}'$. \COMMENT{Low-opacity pruning}
  \ENDIF
  \IF{$\mathbf{area}(b_i) < \tau_{\mathbf{area}}$}
    \STATE Remove $b_i$ from $\mathcal{B}'$. \COMMENT{Small-area pruning}
  \ENDIF
  \IF{$\mathbf{IoU}\!\big(b_i, \{\,b_j \in \mathcal{B} \mid \mathbf{colordiff}(b_i,b_j) < 0.03\,\}\big) > 0.9$}
    \STATE Remove $b_i$ from $\mathcal{B}'$. \COMMENT{Redundant-overlap pruning}
  \ENDIF
\ENDFOR

\STATE \textbf{Densification phase}
\STATE $\mathcal{R} \leftarrow \mathbf{ConnectedComponents}(E)$.
\STATE $\mathcal{R}_{\mathbf{sorted}} \leftarrow \mathbf{SortByArea}(\mathcal{R})$.
\FOR{each region $r_j \in \mathcal{R}_{\mathbf{sorted}}$}
  \IF{curve budget allows}
    \STATE Insert a new closed Bézier curve into $r_j$.
  \ENDIF
\ENDFOR

\STATE \textbf{return} $\mathcal{B}'$.
\end{algorithmic}
\end{algorithm}

% \clearpage

\section{Convert Bézier Splatting to Standard SVG Format}

As described in Algorithm~\ref{alg:svg-export}, we convert our optimized Bézier Splatting representation into a standard SVG file for compatibility with downstream vector editing tools.
Each curve is represented by a set of $3k + 1$ control points $\mathcal{C}_i \in \mathbb{R}^{(3k+1) \times 2}$, corresponding to $k$ continuous cubic Bézier segments.
These control points are normalized to the range $[-1, 1]$ and are first transformed into pixel coordinates according to the canvas size $(W, H)$.

For each curve $i$, we initialize the SVG path string $d$ with a \texttt{[M $P_i^0$]} command.
We then iteratively construct each cubic segment using three internal control points:
$(p_1, p_2, p_3) = (P_i^{3j+1}, P_i^{3j+2}, P_i^{3j+3})$,
and append the segment as a path command \texttt{[C $p_1$, $p_2$, $p_3$]} to $d$.
After all segments are processed, a \texttt{[Z]} is appended to close the path.

Each path $d$ is then associated with a fill color $(r, g, b)$, computed via sigmoid activation on the optimized feature vector $\mathcal{F}_i$, and opacity 1.0.
The result is stored in a global path set $\mathcal{S}$, and all paths are assembled into the final SVG file $S$.
This output is directly compatible with standard vector graphics tools such as Adobe Illustrator.

\begin{algorithm}[!h]
   \caption{\textbf{Convert Bézier Splatting to Standard SVG}}
   \label{alg:svg-export}
\begin{algorithmic}
   \STATE {\bfseries Input:} Optimized control points $\mathcal{C} \in \mathbb{R}^{N \times (3k+1) \times 2}$, feature colors $\mathcal{F} \in \mathbb{R}^{N \times 3}$, canvas size $(W, H)$
   \STATE {\bfseries Output:} SVG file $S$ containing cubic Bézier paths \\
   \FOR{each curve $i = 1$ to $N$}
   \STATE $P_i \leftarrow \frac{\mathcal{C}_i + 1}{2} \cdot (W, H)$
    \STATE $(r, g, b) \leftarrow \text{sigmoid}(\mathcal{F}_i) \times 255$
    \STATE Initialize path $d \leftarrow \texttt{[M $P_i^0$]}$

    \FOR{each segment $j = 1$ to $k$} 
    \STATE $(p_1, p_2, p_3) \leftarrow $($P_i^{3j + 1}, P_i^{3j + 2}, P_i^{3j + 3})$\;
    \STATE $d \oplus \texttt{[C $p_1$, $p_2$, $p_3$]}$ \\
    \ENDFOR
    \STATE $d \oplus \texttt{Z}$\;
    \STATE $\mathcal{S} \leftarrow \mathcal{S} \cup \{ \text{path}(d,\, \text{fill} = (r, g, b),\, \text{opacity} = 1.0) \}$
    \ENDFOR
   \STATE {\bfseries Return:} SVG file assembled from all paths
\end{algorithmic}
\end{algorithm}

\section{Bézier Splatting Supports Flexible Curve Attributes}
All 2D Gaussians in our method are generated via a differentiable sampling algorithm, which makes it straightforward to incorporate user-defined shape attributes, such as linear-gradient fills in color or opacity. To demonstrate the extensibility of our model, we evaluate the multi-opacity scheme for open curves: each Bézier segment within an open curve is assigned an independent opacity value. This design enables users to either maintain consistent sampling point appearance across segments or apply customized interpolation strategies. This strategy significantly improves vectorization quality for open curves, as shown in Table ~\ref{tab:ablation_table}. Since closed curves are the primary representation format in SVG, we adopt a uniform setting for opacity and color in those regions to ensure a full compatibility with existing vector graphic tools. Nonetheless, users can still follow the above multi-opacity scheme to implement linear-gradient fills for both opacity and color, enhancing the flexibility and expressiveness of vector graphics.

\begin{table}[!h]
\centering
\caption{An ablation study on the number of opacity parameters per curve, evaluated on 512 open curves with DIV2K dataset \cite{Timofte_2017_CVPR_Workshops}.}
\vspace{-.5em}
\resizebox{0.45\linewidth}{!}{%
\begin{tabular}{l|ccc}
\hline
  \textbf{Method}  & $SSIM^\uparrow$   & $PSNR^\uparrow$    & $LPIPS^\downarrow$   \\ 
  \hline
  % Full (w/ Prune\&Densify, 3-opacity)  &\textbf{0.65}  &\textbf{23.79}   &\textbf{0.50}  \\
    3-opacity  &\textbf{0.65}  &\textbf{23.79}   &\textbf{0.50}  \\
    % ~~- w/o Prune\&Densify   & 0.63   & 23.23   & \textbf{0.50}   \\
   1-opacity   &0.63    &22.97   &0.51    \\
  % & \checkmark  & 3  &   &  &   \\
  \hline

\end{tabular}
}
\label{tab:ablation_table}
\end{table}

\section{More Comparisons}

% We evaluate the differentiable VG methods on more natural images. We quantitatively evaluate our method on another natural image dataset, Kodak \cite{kodak1999}, and the results are shown in Table \ref{tab:comparisons_kodak} and Fig. \ref{fig:appendix_Kodak}.
% Fig. \ref{fig:app_comparsion_0}, \ref{fig:app_comparsion_1}, \ref{fig:app_comparsion_2}, \ref{fig:app_comparsion_3}, and \ref{fig:app_comparsion_4} show more qualitative comparisons. In the second image, our method successfully captures finer details in the blanket, preserving intricate textures and subtle variations that are less accurately represented by other methods. This demonstrates the superior representation capacity of our Bézier Splatting, particularly for complex textures and high-frequency details while maintaining structural integrity. 

\begin{figure*}[t]
\centering
\includegraphics[width=1.0\linewidth]{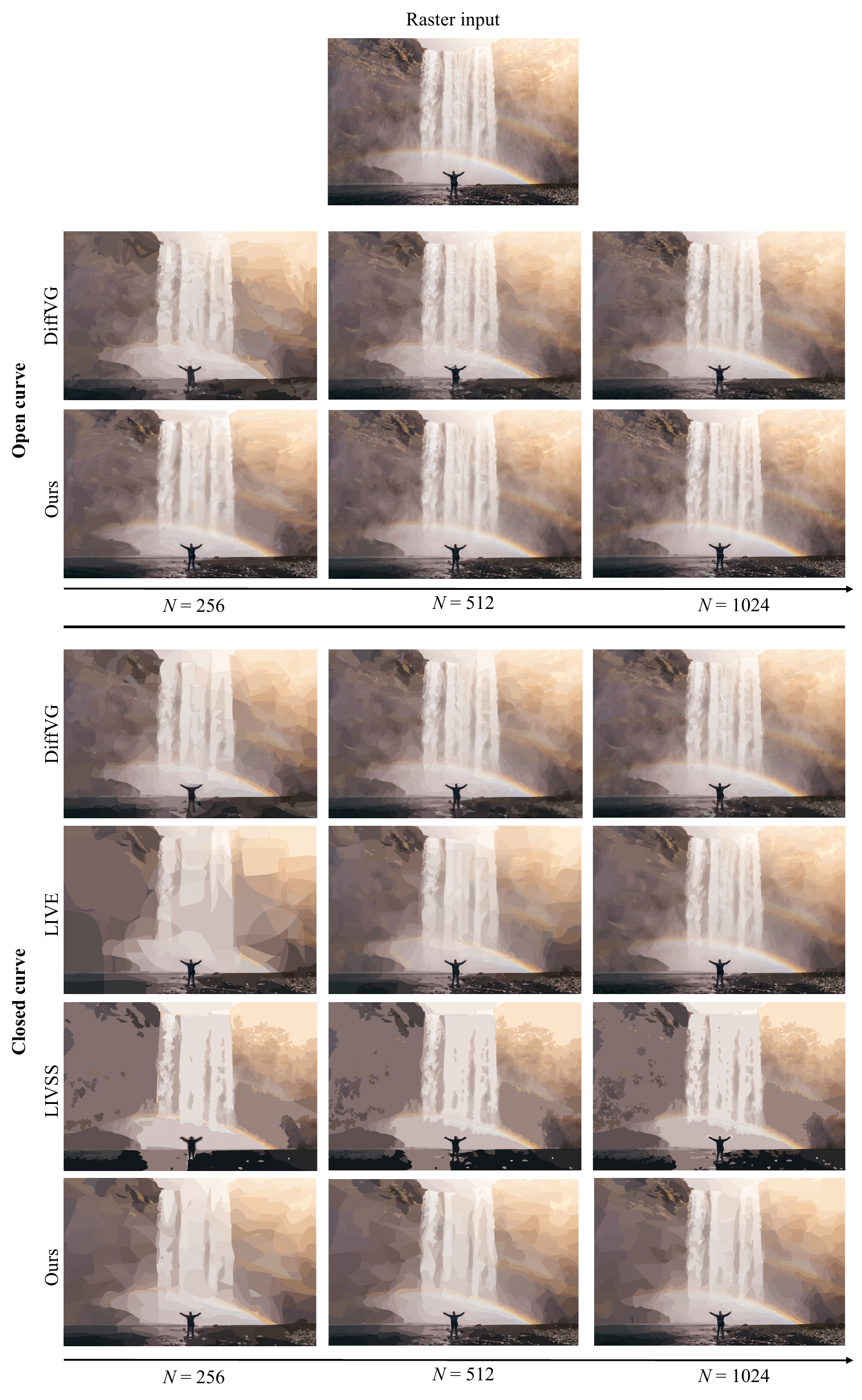}
% \vspace{-2em}
\caption{A qualitative comparison of our method and the existing differentiable VG rasterization method on DIV2K dataset \cite{Timofte_2017_CVPR_Workshops}.
}
\label{fig:app_comparsion_0}
\end{figure*}

\begin{figure*}[t]
\centering
\includegraphics[width=1\linewidth]{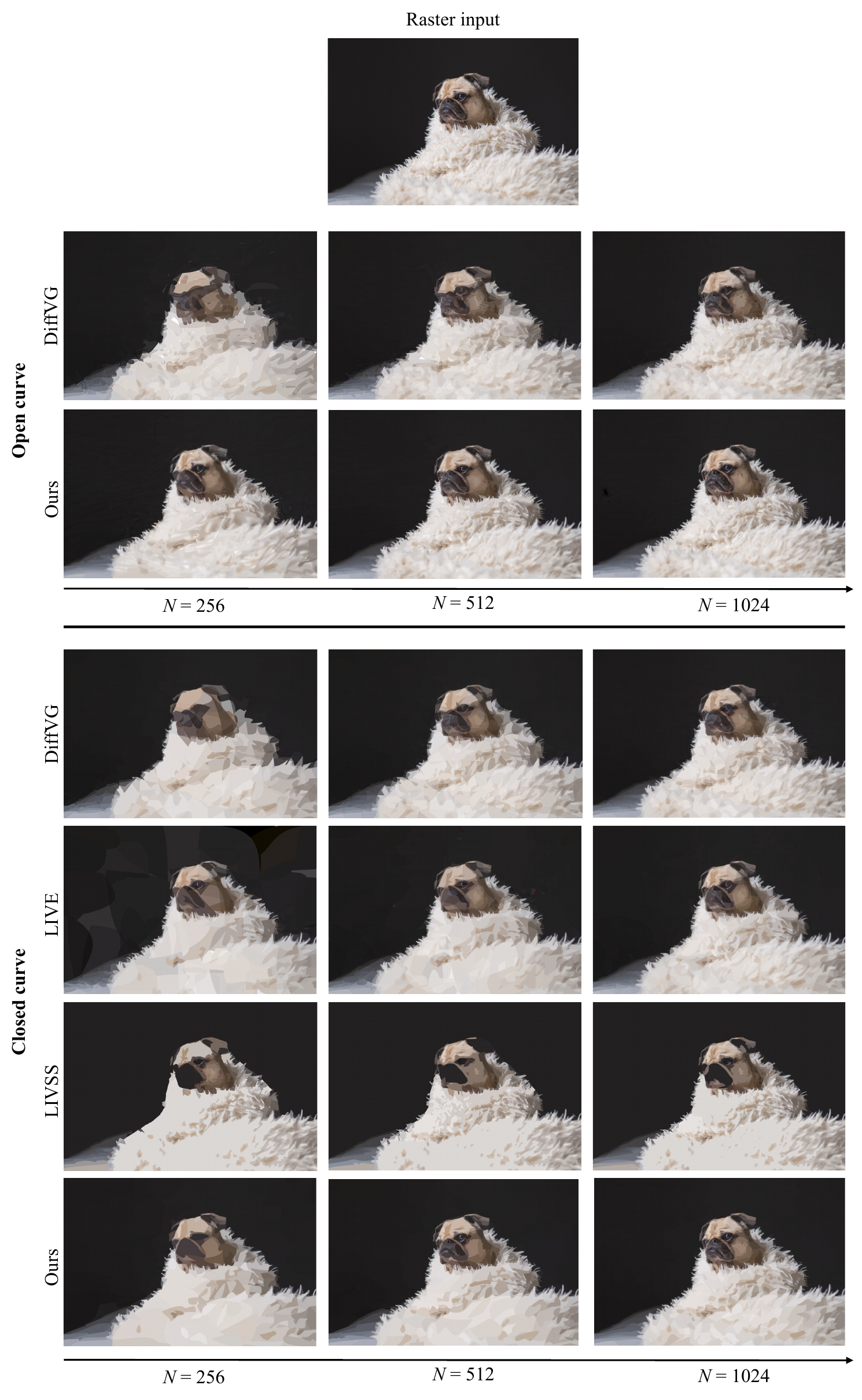}
% \vspace{-2em}
\caption{More qualitative comparisons of our method and the existing differentiable VG rasterization method on DIV2K dataset \cite{Timofte_2017_CVPR_Workshops}.
}
\label{fig:app_comparsion_1}
\end{figure*}

\begin{figure*}[t]
\centering
\includegraphics[width=1\linewidth]{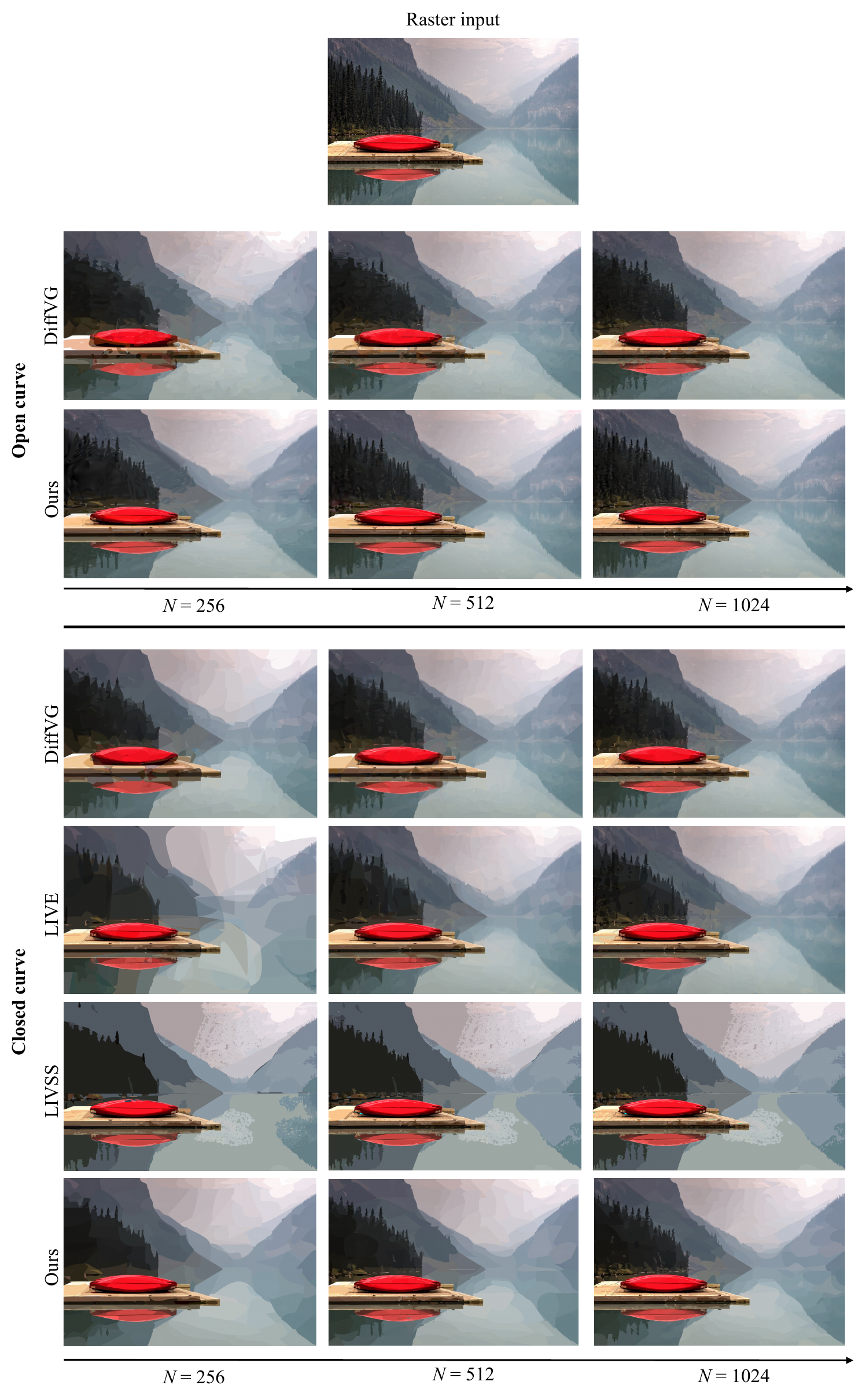}
% \vspace{-2em}
\caption{More qualitative comparisons of our method and the existing differentiable VG rasterization method on DIV2K dataset \cite{Timofte_2017_CVPR_Workshops}.
}
\label{fig:app_comparsion_2}
\end{figure*}

\begin{figure*}[t]
\centering
\includegraphics[width=1\linewidth]{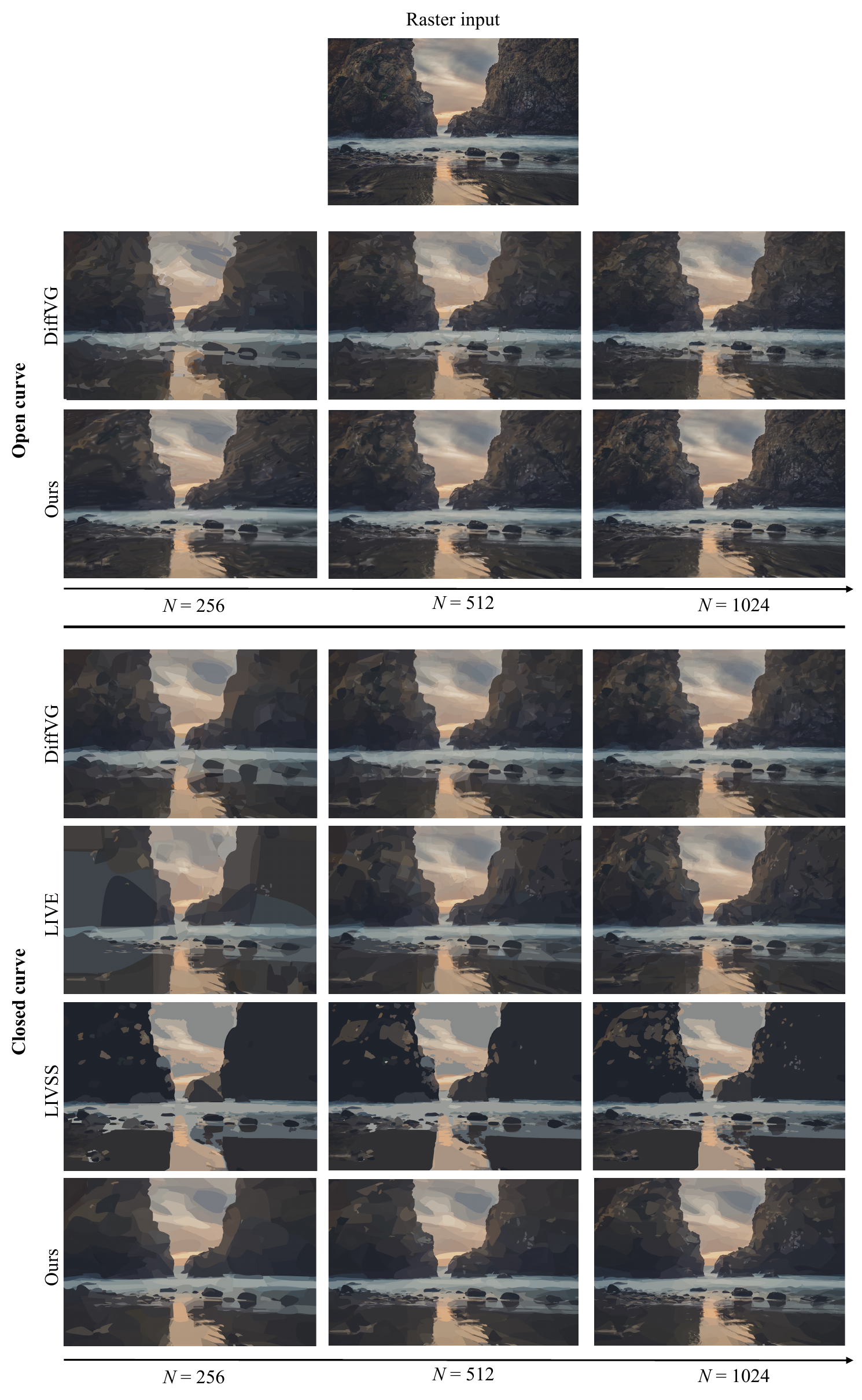}
% \vspace{-2em}
\caption{More qualitative comparisons of our method and the existing differentiable VG rasterization method on DIV2K dataset \cite{Timofte_2017_CVPR_Workshops}.
}
\label{fig:app_comparsion_3}
\end{figure*}

\begin{figure*}[t]
\centering
\includegraphics[width=1\linewidth]{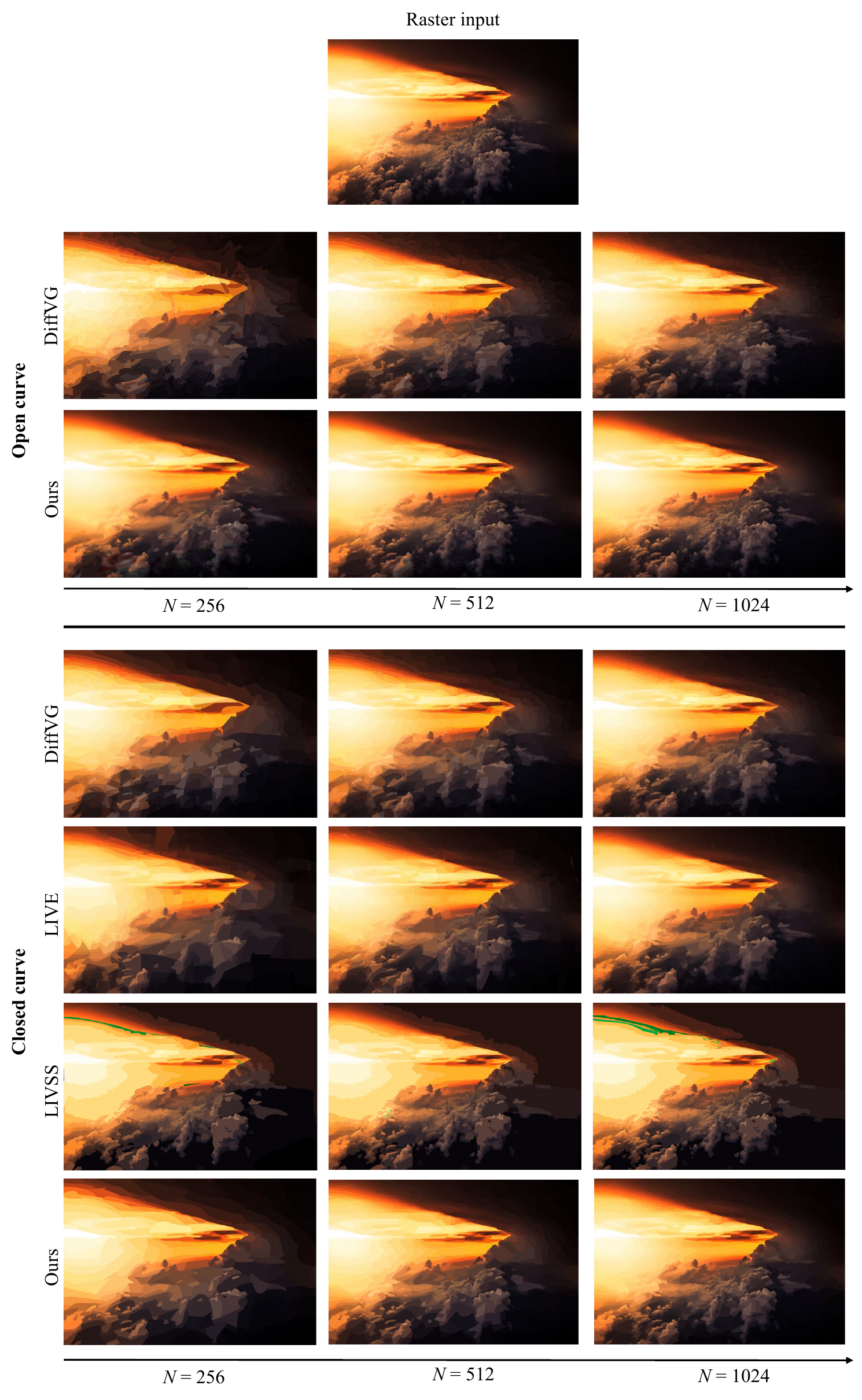}
% \vspace{-2em}
\caption{More qualitative comparisons of our method and the existing differentiable VG rasterization method on DIV2K dataset \cite{Timofte_2017_CVPR_Workshops}.
}
\label{fig:app_comparsion_4}
\end{figure*}

We present more qualitative comparisons on the DIV2K~\cite{Timofte_2017_CVPR_Workshops} dataset. As shown in Fig.~\ref{fig:app_comparsion_0}, Fig.~\ref{fig:app_comparsion_1}, Fig.~\ref{fig:app_comparsion_2}, Fig.~\ref{fig:app_comparsion_3}, and Fig.~\ref{fig:app_comparsion_4}, our method shows consistent improvements over the baselines, including better preservation of fine details, higher rendering fidelity, fewer visual artifacts, and more accurate geometric structures. Furthermore, as the number of curves increases, our method demonstrates a significantly improved representational ability, enabling more precise reconstruction of complex shapes, and fine-grained structures.

We further evaluate our method on another natural image dataset, Kodak~\cite{kodak1999}. Due to the slow vectorization speed of LIVE \cite{xu2022live}, we compare with DiffVG on open curves only. As shown in Table~\ref{tab:comparisons_kodak}, our method consistently outperforms DiffVG across quantitative metrics including PSNR, SSIM, and LPIPS. 

\begin{table*}[!h]
    \centering
    \caption{
        A quantitative evaluation on the Kodak dataset~\cite{kodak1999} with 256 to 1024 curves. 
        We report results on open curves.
        % using SSIM, PSNR (higher is better), and LPIPS (lower is better).
    }
    \resizebox{1\textwidth}{!}{
        \begin{tabular}{l|l|ccc|ccc|ccc}
            \toprule
            \multirow{2}{*} & \multirow{2}{*}{\textbf{Method}} 
             & \multicolumn{3}{c|}{\textbf{256}} 
             & \multicolumn{3}{c|}{\textbf{512}} 
             & \multicolumn{3}{c}{\textbf{1024}} 
              \\
             
             & & $SSIM^\uparrow$ & $PSNR^\uparrow$ & $LPIPS^\downarrow$
               & $SSIM^\uparrow$ & $PSNR^\uparrow$ & $LPIPS^\downarrow$
               & $SSIM^\uparrow$ & $PSNR^\uparrow$ & $LPIPS^\downarrow$ \\
            \midrule

            \multirow{2}{*}{Open} 
            & DiffVG & 0.601 & 23.43 & 0.535  
                    & 0.645 & 24.70 & 0.495  
                    & 0.699 & 26.04 & 0.439   \\
            & Ours   & \textbf{0.679} & \textbf{26.18} & \textbf{0.457}  
                    & \textbf{0.743} & \textbf{27.90} & \textbf{0.383}  
                    & \textbf{0.797} & \textbf{29.24} & \textbf{0.310}  \\

            \midrule
            \multirow{2}{*}{Closed} 
            & DiffVG & \textbf{0.622} & 24.11 & \textbf{0.513}  
                    & \textbf{0.666} & 25.34 & 0.475  
                    & \textbf{0.719} & 26.66 & \textbf{0.420}  \\
            % & LIVE   & \textbf{0.647} & \textbf{24.61} & \textbf{0.495}  
            %         & \textbf{0.693} & \textbf{26.00} & \textbf{0.447}  
            %         & \textbf{0.742} & \textbf{27.47} & \textbf{0.392}  
            %         & - & - & - \\
            & Ours 
                    & 0.621 & \textbf{24.19} & 0.519 
                    & 0.664 & \textbf{25.61} &0.485 
                    & 0.708 & \textbf{26.91} & 0.448  \\
            % % \multirow{3}{*}{Closed} 
            % & DiffVG & 0.622 & 24.11 & 0.513  
            %         & 0.666 & 25.34 & 0.475  
            %         & 0.719 & 26.66 & 0.420  
            %         & 0.779 & 28.17 & 0.350 \\
            % % & LIVE   & \textbf{0.647} & \textbf{24.61} & \textbf{0.495}  
            % %         & \textbf{0.693} & \textbf{26.00} & \textbf{0.447}  
            % %         & \textbf{0.742} & \textbf{27.47} & \textbf{0.392}  
            % %         & - & - & - \\
            % & Ours 
            %         & 0.621 & 24.19 & 0.519 
            %         & 0.664 & 25.61 &0.485 
            %         & 0.708 & 26.91 & 0.448  
            %         & \textbf{0.755} & \textbf{28.02} & \textbf{0.393} \\
            \bottomrule
        \end{tabular}
    }
    \label{tab:comparisons_kodak}
\end{table*}

\section{More Image Vectorization Results}
\begin{figure*}[t]
\centering
\hspace*{-0.205\linewidth} 
\includegraphics[width=1.4\linewidth]{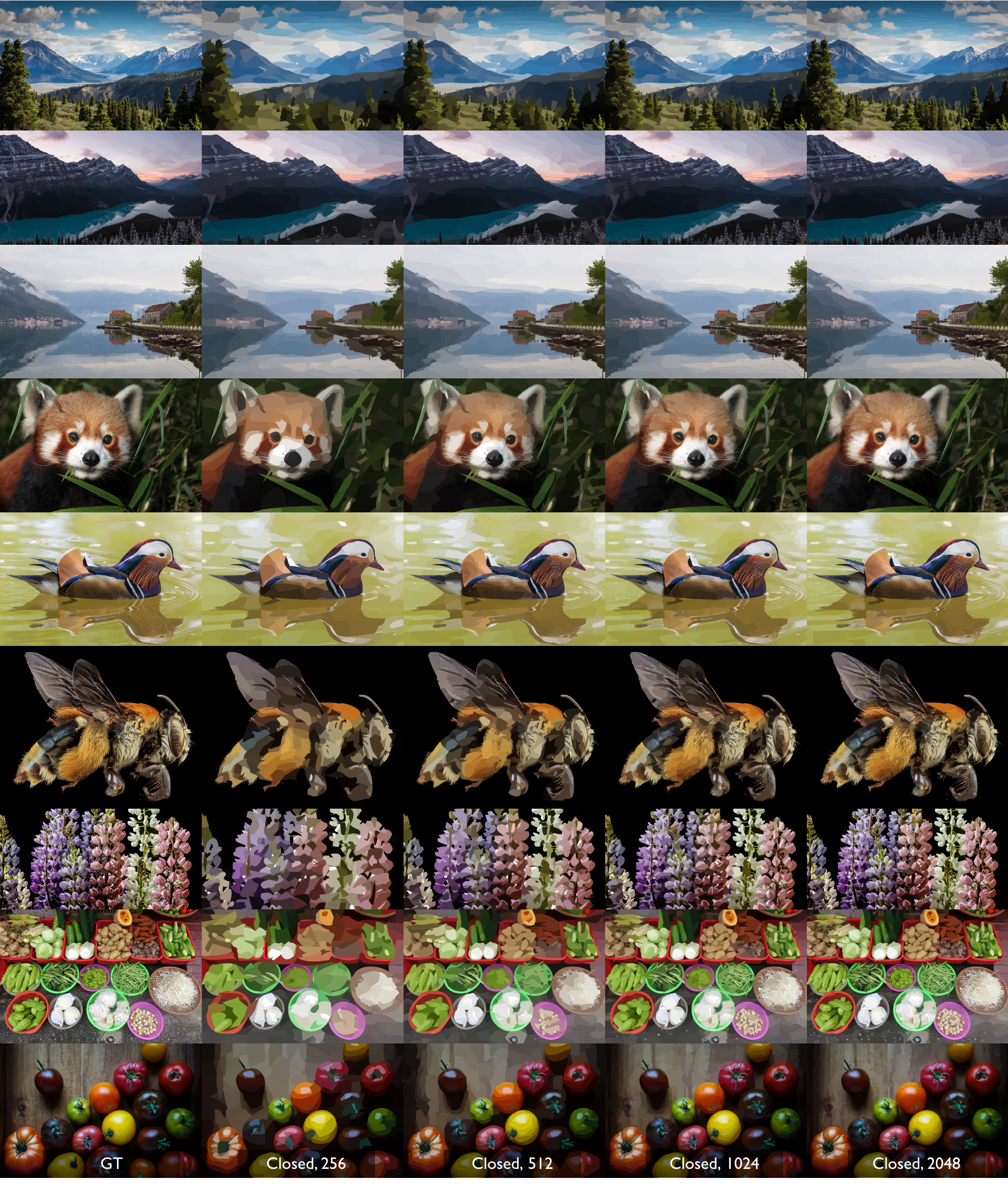}
\caption{More image vectorization results by Bézier Splatting on DIV2K dataset \cite{Timofte_2017_CVPR_Workshops}.}
\label{fig:appendix_large_div}
\end{figure*}

\begin{figure*}[t]
\centering
\hspace*{-0.205\linewidth} 
\includegraphics[width=1.4\linewidth]{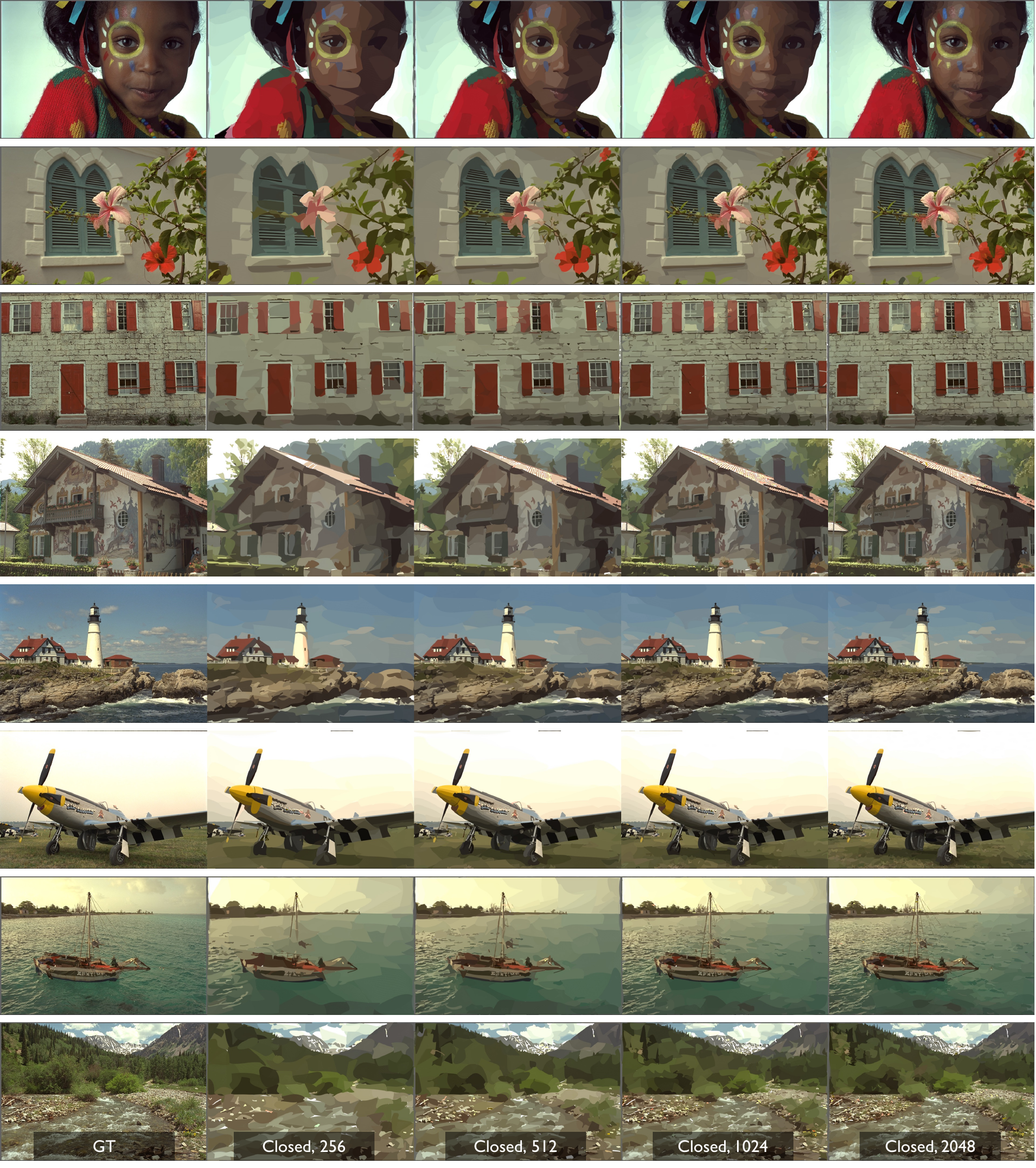}
\caption{More image vectorization results by Bézier Splatting on Kodak \cite{kodak1999} dataset.}
\label{fig:appendix_Kodak}
\end{figure*}

\begin{figure*}[t]
\centering
\hspace*{-0.205\linewidth} 
\includegraphics[width=1.4\linewidth]{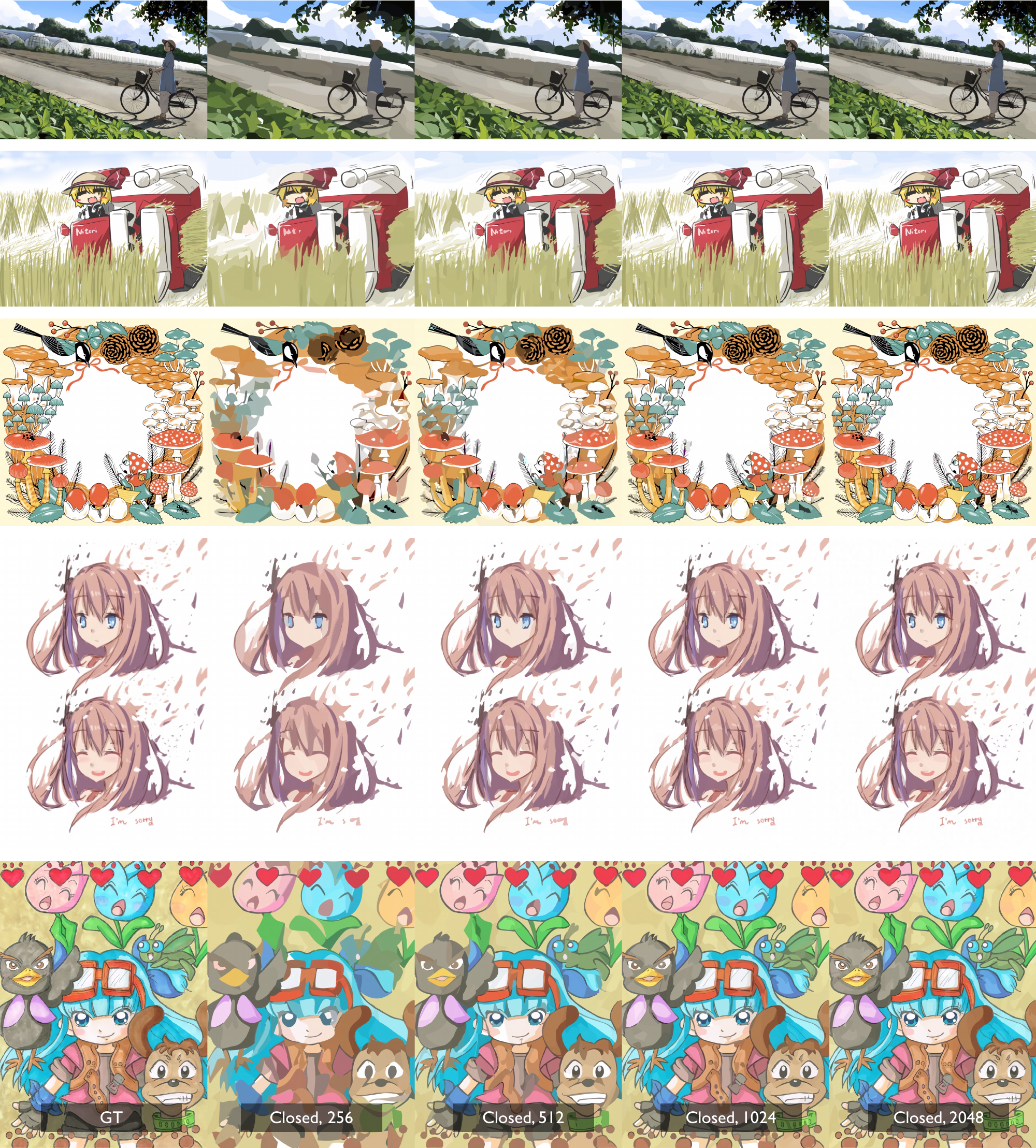}
\caption{More image vectorization results by Bézier Splatting on DanbooRegion dataset \cite{DanbooRegion2020}.}
\label{fig:appendix_large_animation}
\end{figure*}

% We sho our method on more image examples across a broader range of image domains.
% Fig. \ref{fig:appendix_large_line} shows more natural image examples by open curves, while Fig. \ref{fig:appendix_large_closed} shows more results by closed curves.
Fig. \ref{fig:appendix_large_div}, Fig. \ref{fig:appendix_Kodak}, and Fig. \ref{fig:appendix_large_animation} show more results on natural images from DIV2K \cite{Timofte_2017_CVPR_Workshops} and Kodak \cite{kodak1999}, as well as animation images from DanbooRegion \cite{DanbooRegion2020}, respectively. The results demonstrate that both the global structure and local texture of the images are well reconstructed, highlighting the effectiveness of our approach in capturing fine details and complex shapes.

% \begin{figure*}[t]
% \centering
% \includegraphics[width=1\linewidth]{figures/appendix_large_div_line.pdf}
% \caption{More image vectorization results by Bézier Splatting on DIV2K dataset \cite{Timofte_2017_CVPR_Workshops}.}
% \label{fig:appendix_large_line}
% \end{figure*}

% \begin{figure*}[t]
% \centering
% \includegraphics[width=1\linewidth]{figures/appendix_large_div_area.pdf}
% \caption{More image vectorization results by Bézier Splatting on DIV2K dataset \cite{Timofte_2017_CVPR_Workshops}.}
% \label{fig:appendix_large_closed}
% \end{figure*}

% \section{Technical Appendices and Supplementary Material}
% Technical appendices with additional results, figures, graphs and proofs may be submitted with the paper submission before the full submission deadline (see above), or as a separate PDF in the ZIP file below before the supplementary material deadline. There is no page limit for the technical appendices.

%%%%%%%%%%%%%%%%%%%%%%%%%%%%%%%%%%%%%%%%%%%%%%%%%%%%%%%%%%%%
\clearpage

\end{document}